\documentclass[prd,aps,letterpaper,twocolumn,superscriptaddress,preprintnumbers,nofootinbib,floatfix]{revtex4}
\pdfoutput=1
\usepackage{color}
\usepackage[pdftex]{graphicx}
\usepackage{amsmath}
\usepackage{amsfonts}
\usepackage{amssymb}
\usepackage{rotating}
\usepackage{subfigure}
\usepackage{paralist} 
\usepackage{verbatim}
\usepackage{float}
\usepackage{soul} 
\usepackage{appendix}
\usepackage{natbib}
\usepackage{epsfig}
\usepackage[colorlinks=true,linkcolor=red,urlcolor=blue,citecolor=blue]{hyperref}
\usepackage[utf8]{inputenc}
\usepackage[english]{babel}
\usepackage{tabularx}
\usepackage{url}
\usepackage{url}

\newcolumntype{C}{>{\centering\arraybackslash}X}
\usepackage{mathpazo} 
\usepackage{tgpagella} 

\newcommand{\cca}{Center for Computational Astrophysics, Flatiron Institute, 162 5th Avenue, New York, NY, USA 10010}
\newcommand{\princetonAstro}{Department of Astrophysical Sciences, Princeton University, 4 Ivy Lane, Princeton, NJ, USA 08544}
\newcommand{\stonybrook}{Physics and Astronomy Department, Stony Brook University, Stony Brook, NY  11794}

\begin{document}

\title{MillimeterDL: Deep Learning Simulations of the Microwave Sky}
\author{Dongwon Han}
\affiliation{\stonybrook}

\author{Neelima~Sehgal}
\affiliation{\stonybrook}

\author{Francisco Villaescusa-Navarro}
\affiliation{\princetonAstro}
\affiliation{\cca}

\begin{abstract}
We present 500 high-resolution, full-sky millimeter-wave Deep Learning (DL) simulations that include lensed CMB maps and correlated foreground components. 
We find that these MillimeterDL simulations can reproduce a wide range of non-Gaussian summary statistics matching the input training simulations, while only being optimized to match the power spectra.
The procedure we develop in this work enables the capability to mass produce independent full-sky realizations from a single expensive full-sky simulation, when ordinarily the latter would not provide enough training data.  We also circumvent a common limitation of high-resolution DL simulations that they be confined to small sky areas, often due to memory or GPU issues; we do this by developing a ``stitching'' procedure that can faithfully recover the high-order statistics of a full-sky map without discontinuities or repeated features. In addition, since our network takes as input a full-sky lensing convergence map, it can in principle take a full-sky lensing convergence map from any large-scale structure (LSS) simulation and generate the corresponding lensed CMB and correlated foreground components at millimeter wavelengths; this is especially useful in the current era of combining results from both CMB and LSS surveys, which require a common set of simulations.
\end{abstract}

\maketitle

\section{Introduction}
\label{sec:intro}
The Cosmic Microwave Background (CMB) has been a cornerstone of modern precision cosmology. The study of the primordial anisotropy imprinted on the CMB at the surface of last scattering has shed light on the physics of the early Universe. A rich set of secondary anisotropies induced by intervening Large Scale Structures (LSS) and complex astrophysical phenomena has revealed a picture of the evolving Universe. As the sensitivity of CMB experiments has improved, more information has been obtained from these features. A few recent examples include the constraints on cosmological parameters from CMB power spectra ~\cite{Planck2018,Henning2018,Aiola2020,Choi2020,Han2020}, the measurement of CMB lensing auto spectra ~\cite{Sherwin2017, Planck2018Lensing, Wu2019}, and the study of kinematic and thermal Sunyaev-Zel'dovich effects ~\cite{Planck2016SZ, Amodeo2020, Schaan2020, Huang2020}.

Current CMB data analyses rely heavily on simulations to (1)~model the underlying physical processes, (2)~verify the data analysis pipeline, (3)~investigate potential biases, and (4)~generate covariance matrices used in likelihood analyses. As a result, there have been efforts to improve simulations of the microwave sky~\cite{Jaffe1999, Nuevo2005, Schafer2006, Roncarelli2007, Carbone2008, Battaglia2010, Sehgal2010, Flender2016, Takahashi2017, Stein2020}. Current and future high-resolution CMB experiments, such as the Atacama Cosmology Telescope (ACT)~\cite{ACT2007,ACT2016}, the South Pole Telescope (SPT)~\cite{SPT2011}, the Simons Observatory (SO)~\cite{SO2019}, CMB-S4~\cite{S42019}, and CMB-HD~\cite{Sehgal2019, Sehgal2020} will observe large sky areas at arcminute resolution or better with unprecedented sensitivity. Analyzing the data from these observations will need many realizations of realistic simulations of comparable resolution and sky coverage.  In particular, as the sensitivity of CMB experiments improves, simulations capturing the correlations between foreground components and non-Gaussian features will become ever more important.  

Simulations of the millimeter-wave sky can be divided into several components: (1)~primordial CMB maps, (2)~extragalactic foregrounds, and (3)~Galactic foregrounds. The procedure to generate simulations of primordial CMB maps is well-established and computationally efficient at arcminute resolution~\cite{Lewis2005, Amblard2004, Basak2008, Fabbian2013, Naess2013}. The extragalactic foreground components include the lensing convergence field ($\kappa$), the kinetic and thermal Sunyaev-Zel'dovich effects (kSZ and tSZ), the Cosmic Infrared Background (CIB), and radio galaxies (Radio). There are a few extragalactic foreground simulations publicly available that include all the above non-Gaussian, correlated foregrounds at high-resolution over a large fraction of the sky~\cite{Sehgal2010, Stein2020}. 
These simulations are generated by first making three-dimensional realizations of cosmological features, and then projecting them along the line-of-sight to make two-dimensional maps. 
To a large extent, these existing simulations can reproduce the statistics of the latest CMB observations. However, there are only a limited number of realizations available due to the high computational cost of generating initial three-dimensional realizations.  As a result, many CMB analyses model extragalactic foregrounds as Gaussian random fields matched to the observed foreground power spectra (e.g.~\cite{Choi2020, Han2020, Sherwin2017}). 

There have also been a number of efforts to simulate Galactic foregrounds in both intensity and polarization~\cite{Baccigalupi2003, Hennebelle2008, Waelkens2009, Choi2015, Thorne2017, Vansyngel2017, Beck2020, Aylor2021, Thorne2021}.  Modeling the small-scale features of Galactic foregrounds is particularly challenging given the current lack of observations on comparable scales in millimeter wavebands, though there is progress in this direction~\cite{Krachmalnicoff2020}.  In this work, we focus on generating simulations including the primordial CMB and extragalactic foreground components, and leave inclusion of Galactic foregrounds to future work.

Recently, Machine Learning has found a broad application in cosmology; parameter estimation~\cite{Levasseur2017, Ravanbakhsh2017, Gupta2018, Mathuriya2018, Alsing2019, Shirasaki2019,  Lucie-Smith2020, Villaescusa-Navarro2020}, foreground cleaning~\cite{Petroff2020, Aylor2021, Thorne2021, Moriwaki2021}, feature extraction~\cite{Caldeira2019, Bonjean2020, Guzman2021, Lin2021}, inpainting~\cite{Puglisi2020,Montefalcone2020, Sadr2020}, filtering~\cite{Smith2019}, and modeling of physical processes~\cite{Lucie-Smith2018, Lovell2019, Alves2020, Chen2020v2, Cranmer2020, Thorne2021, Wadekar2020, Wang2021}. Deep Learning (DL) is a subset of Machine Learning that fits {\it{multiple}} non-linear functions to the input data using neural networks. In particular, generative DL models aim at sampling the manifold (which can be high-dimensional) where the input data lives, so that new points in that manifold are associated with new complex data with the same statistical properties as the desired input data. Recently, there have been many approaches in this direction~\cite{Berger2019, He2019, Mustafa2019, Perraudin2019, Troster2019,  Chen2020, Curtis2020, Dai2020, Krachmalnicoff2020, Li2020, Tamosiunas2020, Thiele2020, Ullmo2020, Aylor2021, Thorne2021}. However, these methods are not yet widely used in CMB data analyses due to their limited sky-coverage or summary statistics mismatch.

In this work, we generate for the first time, many independent full-sky millimeter-wave DL simulations that include all correlated extragalactic foreground components appropriate for CMB analyses. We find that our MillimeterDL (hereafter mmDL) simulations are able to reproduce well the non-Gaussian statistics of the original input maps.  Several recent works have focused on using various DL techniques to generate two-dimensional projected microwave fields from three-dimensional N-body simulations~\cite{Berger2019, He2019, Perraudin2019, Troster2019,  Chen2020, Curtis2020, Dai2020, Li2020, Thiele2020, Ullmo2020}; in this work, we use two-dimensional projected simulations at 148~GHz described in~\cite{Sehgal2010} (hereafter S10) as the input data to train our DL algorithm, as opposed to three-dimensional N-body simulations. 
This is suitable for generating maps for a CMB imaging survey, since most of the relevant information for CMB data analyses is contained in the two-dimensional projected maps. The advantage of our approach is that DL training in two dimensions is computationally more efficient than that in three dimensions; therefore, for a given computing resource, one can train on a larger footprint at higher-resolution. 

In particular, we use DL methods to generate 500 realizations of full-sky millimeter-wave simulations that match the statistics of the S10 simulations. Each of these 500 simulation realizations includes:
\begin{itemize}
    \item a full-sky simulation at half-arcminute resolution at six different frequencies (30, 90, 148, 219, 277, and 350~GHz) 
    
    \item the lensed CMB in both temperature and polarization
    \item extragalactic foreground components appropriately correlated with each other, including: 
    \begin{itemize}
    \item the lensing convergence map ($\kappa$)
    \item the kinetic  Sunyaev-Zel'dovich effect (kSZ)
    \item the thermal Sunyaev-Zel'dovich effect (tSZ)
    \item the Cosmic Infrared Background (CIB)
    \item the radio galaxies (Radio)
    \end{itemize}
    \item non-Gaussian information that reproduces statistics such as the one-point function, bispectra, and trispectra.  
\end{itemize}

These mmDL simulations are trained using the cosmology and extragalactic model described in S10~\cite{Sehgal2010}. However, the method presented here is general and can be applied to reproduce other two-dimensional simulations with minor modifications. In Section~\ref{sec:method}, we present our procedure to generate the mmDL simulations. The comparison between the statistical properties of the mmDL and the S10 simulations is shown in Section~\ref{sec:results}. We discuss potential applications and the data products released in Section~\ref{sec:discussion}.  These mmDL simulations are publicly available in both HEALPix{\color{blue}\footnote{\href{http://healpix.sourceforge.net}{http://healpix.sourceforge.net}\label{fn:healpix}}} and CAR formats on the Legacy Archive for Microwave Background Data Analysis (LAMDA){\color{blue}\footnote{\href{https://lambda.gsfc.nasa.gov/simulation/tb_sim_ov.cfm}{https://lambda.gsfc.nasa.gov/simulation/tb\_sim\_ov.cfm}\label{fn:lambda}}} and the National Energy Research Scientific Computing Center (NERSC) cluster{\color{blue}\footnote{\href{https://crd.lbl.gov/departments/computational-science/c3/c3-research/cosmic-microwave-background/cmb-data-analysis-at-nersc/}{https://crd.lbl.gov/departments/computational-science/c3/c3-research/cosmic-microwave-background/cmb-data-analysis-at-nersc/}\label{fn:nersc}}}.  We also provide the code to produce additional realizations at~\href{https://github.com/dwhan89/cosmikyu}{https://github.com/dwhan89/cosmikyu}.

\section{DL Primary Input Data}
\label{sec:dataset}
We generate our {\it{Primary Input Data}} from the S10 simulations, which include $\kappa$, kSZ, tSZ, CIB and Radio~\cite{Sehgal2010} extragalactic foreground components. The S10 simulations were generated by post-processing the output of a three-dimensional simulation, which has {\it{WMAP5}} cosmology~\cite{WMAP2009}.  The S10 simulations have correlated foreground components, and have been shown to have non-Gaussian statistics that match observations~\cite{Lacasa2012, Lacasa2013, Hand2012, Hill2013, vanEngelen2014}. The S10 simulations and source catalogs are publicly available on NASA's LAMBDA website (see Footnote~\ref{fn:lambda}). The simulations are saved in HEALPix format~\cite{Gorski2005} with $\rm{Nside}=8192$, and have units of Jy/sr. The S10 $\kappa$ map has $\rm{Nside}=4096$, and is dimensionless. 

We describe our procedure to prepare the mmDL Primary Input data below. Our procedure includes (1)~pre-processing the original S10 148~GHz simulations, (2)~dividing the pre-processed S10 simulations into small patches, and (3)~applying a normalization to the data sets to stabilize and speed up the training of a neural network.

\subsection{Pre-processing of the S10 Simulations}
\label{sec:prep1}

Below we detail the steps in our procedure to pre-process the original HEALPix S10 simulations for the network training.

\begin{itemize}

    \item Following~\cite{vanEngelen2014, SO2019}, we scale the flux of CIB sources and the tSZ map by a factor of 0.75 to make the S10 simulations more closely match recent measurements from ACT, Planck and SPT~\cite{Dunkley2013, Sievers2013, Planck2014, George2015, Planck2016tSZ}. 

    \item We convert the units of the maps from Jy/sr to $\mu$K, except for the $\kappa$ map which is dimensionless. 
    
    \item We reproject the $\kappa$, tSZ, and kSZ foreground components of the S10 simulations from HEALPix pixelization to plate carrée (CAR) pixelization at half arcminute resolution using the public {\it pixell}{\color{blue}\footnote{\href{https://github.com/simonsobs/pixell}{https://github.com/simonsobs/pixell}}} and {\it libsharp}{\color{blue}\footnote{\href{https://github.com/Libsharp/libsharp}{https://github.com/Libsharp/libsharp}}}~\cite{Reinecke2013} libraries. Since CAR maps intrinsically consist of a grid of rectangular pixels, there is a natural way to extract two-dimensional submaps; this makes it simpler to train DL networks and apply two-dimensional Fourier transform operations.  On the other hand, CAR pixels are increasingly over-sampled farther away from the equator (the image is stretched), which we discuss below.  The reprojection is done by converting the original HEALPix maps to spherical harmonic $a_{lm}$ up to an $\ell_{max}$ of 10,000. These $a_{lm}$ are then reprojected onto full-sky CAR pixelized maps.
    
    \item For the CIB and Radio foreground components, we place sources directly from the S10 catalogs into full-sky CAR pixelized maps to avoid spectral leakage ("ringing") around point-like sources when manipulating them in harmonic space.  We apply a flux-cut of 7~mJy at 148~GHz to both CIB and Radio CAR maps in order to match current and future CMB analyses that typically detect and remove bright sources. 

    \item Since CAR maps require the same number of pixels at each declination, the image appears stretched at the poles.  To prevent the network from learning non-physical features induced by this stretching, we use only the region of the simulations near the equator to train the network (from -$10^{\circ}$ to $10^{\circ}$ in declination). However, in order to make use of all the sky area available in the S10 simulations, we rotate the full S10 map many times, redefining the equator of the map each time. In particular, we rotate each full-sky map by 15 degrees in $\psi$ and 20 degrees in $\theta$, where $\psi$ and $\theta$ are the first two Euler angles. The successive rotations in $\psi$ and $\theta$ described above generate 25 unique cylindrical strips around each new equator, which strips in CAR projection have minimal image distortion compared to the sphere. For $\kappa$, tSZ, and kSZ, the rotations are done in spherical harmonic space. For CIB and Radio, we simply remap source positions for given rotation angles.
\end{itemize}

\subsection{Division of Full Sky Simulations into Smaller Patches}
\label{sec:prep2}
To train a network, we need many training samples. Since we start with a single realization of the full sky S10 simulation, we increase the number of samples by dividing this simulation into smaller overlapping patches. Below we detail our steps. 

\begin{itemize}
    \item We randomly pull out patches of $128\times 128$ pixels (roughly $1^{\circ}$ x $1^{\circ}$ patches) between the declination of $-10^{\circ}$ and $10^{\circ}$.  We cut out patches in this way from our 25 cylindrical strips (described above) for $\kappa$, tSZ, kSZ, CIB and Radio maps.  Thus each patch has five foreground components totaling an array of size $5\times 128\times 128$.  We repeat this step until we collect 30,000 validation samples (to pick initial parameters for the training), 200,000 training samples, and 30,000 test samples (to check the performance of the training). We ensure that we have an equal number of samples from the 25 rotated maps. We call the resulting data set the Primary Input Data. Note that these samples will necessarily have overlapping regions among them. We discuss our method to overcome this issue in Section~\ref{sec:method}. 
    
    \item  During the network training, we randomly flip a sample either vertically or horizontally, or both as a natural way to augment the training data set.
\end{itemize}

\subsection{Normalization to Speed up and Stabilize the Training}
\label{sec:prep3}

We need to normalize the samples to stabilize and to speed up the neural network training. Since we have multiple foreground components in a single sample, the normalization needs to account for two issues. First, the pixel intensity Probability Density Function (pixel PDF) of different foreground components follows different distributions. The pixel PDFs of tSZ, CIB, and Radio roughly follow a Poisson distribution, whereas those of $\kappa$ and kSZ are better described by a Gaussian distribution. Second, these pixel PDFs have varying dynamic ranges. For example, all the pixel values of the $\kappa$ map fall between -1.5 and 1.5; on the other hand, the Radio map has pixel values ranging from 0 to 852 $\mu$K. Ideally, we would like the pixel PDFs to follow a standard Gaussian distribution (i.e.~with unit variance) for faster and more stabile network training.  To address the shape of the pixel PDF, we first perform a pixel-wise natural log scaling adopted from~\cite{Troster2019}, to the tSZ, CIB, and Radio maps.  In particular, we replace the value of each pixel, given by $v$, with 

\begin{equation}
    f(v) = {\rm{sgn}}(v)\, \ln \left[\, \frac{{\rm{abs}}(v)}{\sigma_{v}}+1\,\right]\label{eqn:scaled_log_norm}
\end{equation}
where ${\rm{sgn}}(v)$ is sign function, and $\sigma_{v}$ the standard deviation of the pixel PDF of the input map. After this normalization, the pixel PDFs of the tSZ, CIB, and Radio maps are closer to a Gaussian distribution.

To mitigate the dynamic range issue, we then standardize the maps using the standard score method (i.e.~subtract the mean of all the pixels from each pixel value and then divide that by the variance of the map).{\color{blue}\footnote{Another popular approach is to scale an image such that all the pixel values lie between -1 and 1. During our network training, we found that using the the standard score method leads to more stable training compared to this rescaling.}}

\begin{equation}
    s(v') =( v^{'} - {\mu_{v^{'}}} ) / \sigma_{v^{'}} \label{eqn:standerization}   
\end{equation}
Here, $v'$ is the pixel value, and $\mu_{v^{'}}$ and $\sigma_{v^{'}}$ are the mean and standard deviation of the pixel values respectively.{\color{blue}\footnote{Note that we distinguish $v$ and $v'$ here because the pixel values of the tSZ, CIB, and Radio maps are first scaled by Equation~\ref{eqn:scaled_log_norm} (i.e. $p'=f(p)$), whereas $p'=p$ for the $\kappa$ and kSZ maps.}} Note that each foreground component is normalized independently from the others. The parameter values used in the normalization procedure are shown in Appendix~\ref{sec:norm}. Note that each of these normalization operations has a well-defined inverse operation. When we generate the final  products, we undo these normalizations prior to the interpolation procedures discussed in Section~\ref{sec:postprocess}.

\begin{figure*}[t]
  \centering
  \hspace{-3mm}\includegraphics[width=\textwidth]{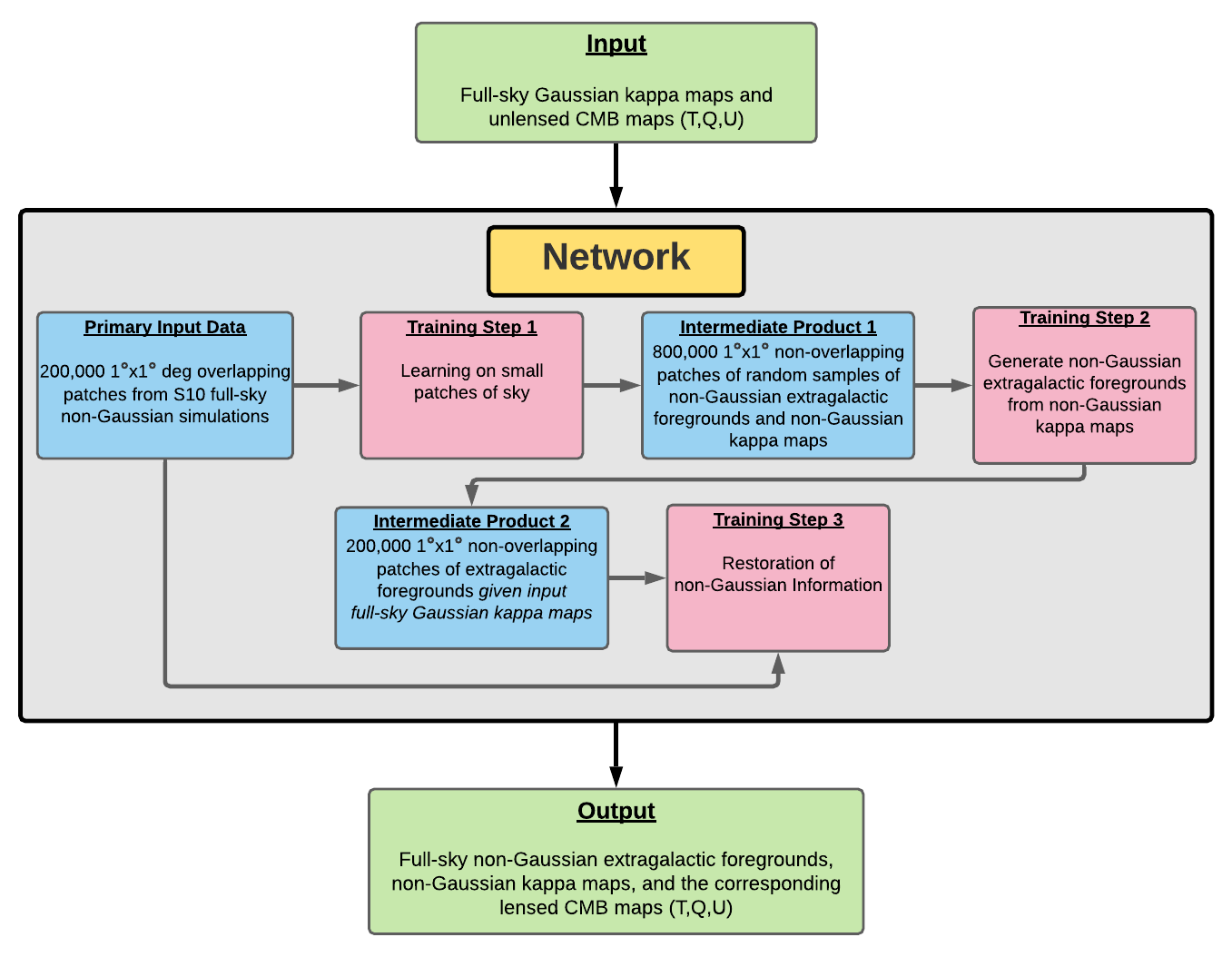}
  \caption{Shown is a schematic of the overall procedure to train the network and generate the output maps from the input data. The network is represented by the shaded gray box in the center. The mmDL procedure starts with the Primary Input Data consisting of 200,000 $1^{\circ}$ x $1^{\circ}$ overlapping patches cut out from the original S10 simulations. The pink boxes represent the three major training steps described in Section~\ref{sec:method}.  The blue boxes represent inputs into each training step. The dark grey arrows indicate the input and output of each step. Once the network training is completed, we feed full-sky Gaussian kappa maps and unlensed CMB maps ($T,Q,U$) into the network (top green box); given these inputs, the network generates the final output products (bottom green box), which are full-sky millimeter-wave simulations including lensed $T, Q$, and $U$ maps, non-Gaussian kappa maps, and non-Gaussian extragalactic foregrounds correlated with the kappa map and each other.}
  \label{fig:flowchart}
  \vspace{2mm}
\end{figure*}

\section{Method}
\label{sec:method}

Generative Adversarial Networks (GANs) are a class of networks that will generate new data matching the statistics of the training data set~\cite{Goodfellow2014}. Fundamental to a GAN is a set of two neural networks, called the generator and discriminator. A generator makes new data, while a discriminator tries to distinguish between the real data and the generated data. By iteratively training a generator and a discriminator, a GAN reaches an equilibrium where the discriminator can no longer distinguish new data from training data. For more discussion about GANs, we refer the reader to~\cite{Goodfellow2017, Creswell2018}.

GAN models can only generate random samples limited to the same footprint as the training data. Since our Primary Input Data consists of $1^{\circ}$ x $1^{\circ}$ patches, the random samples output by the GAN do not have the full information about modes on scales larger than 1 degree (roughly $\ell \le 200$).  This presents a challenge in stitching together the small patches to make a full-sky simulation with the correct large-scale fluctuations.  Furthermore, there is no straightforward way to tile these random patches together to make a full-sky map without having discontinuities at the tile edges.

Instead, we train a conditional GAN network to predict the four extragalactic foreground components (tSZ, kSZ, CIB, and Radio) from a $\kappa$ map. Once this network training step is completed, we generate a full-sky Gaussian $\kappa$ map, and convert it to the other four full-sky extragalactic components tile by tile. Since the input Gaussian $\kappa$ maps is continuous, the resulting extragalactic foreground maps are also continuous. This image prediction approach also solves another major concern about using DL to generate new samples, which is mode collapse. Mode collapse is when the network predicts only a subset of all possible outcomes, instead of a continuous distribution. However, with our approach, since we provide a unique input full-sky $\kappa$ map to the network for each full-sky realization the network outputs, we can ensure that the output maps are also unique.

A challenge of this predictive approach, which uses a conditional GAN model, is that it has a large number of parameters that need to be trained; thus this conditional GAN requires a much larger training data set than simple GANs. Since our Primary Input Data is relatively small and has over-lapping patches (see Section~\ref{sec:dataset} for details), we first use the Primary Input Data to train a simple GAN network (which is easier to train) to make a large sample of non over-lapping patches of $\kappa$, tSZ, kSZ, CIB and Radio maps (Intermediate Product 1 in Figure~\ref{fig:flowchart}). This larger data set is then used to train a conditional GAN (Training Step 2 in Figure~\ref{fig:flowchart}).  After Training Step 2 is done, the network can now predict extragalactic foregrounds given a $\kappa$ map. We input a unique full-sky Gaussian $\kappa$ map into the network to generate the other extragalactic foreground components (tSZ, kSZ, CIB and Radio) tile by tile (Intermediate Product 2 in Figure~\ref{fig:flowchart}). The extragalactic components predicted from the Gaussian $\kappa$ map will be missing some non-Gaussian information due to the missing non-Gaussian information in the $\kappa$ map; thus, we introduce a simpler version of the conditional GAN that can restore this missing non-Gaussian information, which we train on the Primary Input Data plus Intermediate Product 2. Once this training is done, the network is equipped to go from unique input full-sky Gaussian $\kappa$ maps and unlensed CMB $T, Q,$ and $U$ maps to a full-sky simulation including non-Gaussian extragalactic foregrounds, a non-Gaussian $\kappa$ map, and the corresponding lensed CMB $T, Q,$ and $U$ maps ($T$, $Q$, and $U$ are the temperature and two polarization fields of the CMB).  The schematic of the overall procedure is shown in Figure~\ref{fig:flowchart}.  

Since the extragalactic maps are statistically invariant under translation, a natural choice is to use a deep convolutional generative adversarial network (DCGAN) since it has convolution kernels that are also invariant under translation.  In this work, we use three variations of the DCGAN: 1)~Deep Convolutional Wasserstein GAN with gradient penalty (DCWGAN-GP) to make Intermediate Product 1 (this is Training Step 1 discussed in Section~\ref{sec:wgan-gp}), 2)~Pix2Pix GAN (PIXGAN) to make to make Intermediate Product 2 (this is Training Step 2 discussed in Section~\ref{sec:pix2pix}), and 3)~Variational Auto-Encoder GAN (VAEGAN) to augment the missing non-Gaussian information (this is Training Step 3 discussed in Section~\ref{sec:vaegan}).

\subsection{Training Step 1: Learning on Small Patches with Wasserstein GAN with Gradient Penalty (DCWGAN-GP)}
\label{sec:wgan-gp}
\begin{figure*}[t]
  \centering
  \hspace{-3mm}\includegraphics[width=0.8\textwidth]{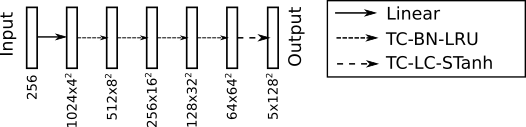}
  \caption{Shown is a diagram of the DCWGAN-GP generator model described in Section~\ref{sec:wgan-gp} and indicated in the schematic of Figure~\ref{fig:flowchart} as Training Step 1. The input to this generator is a latent-vector consisting of 256 random numbers drawn from a Gaussian distribution with mean equal to zero and variance equal to one ($\mu=0$, $\sigma=1$).  The output is one set of $1^{\circ}$ x $1^{\circ}$ maps consisting of the five foreground components. The direction of data flow is indicated by arrows. A solid arrow is a fully connected Linear layer. A short-dashed arrow, labeled TC-BN-LRU, represents a sequence of 1)~a $4\times 4$ two-dimensional transposed convolution, 2)~a batch normalization, and 3)~a leaky rectified linear unit activation function with $\alpha=0.2$. Lastly, a long-dashed arrow, labeled TC-LC-STanh, is a sequence of 1)~a $4\times 4$ two-dimensional transposed convolution, 2)~a Linear Channel, and 3)~a Scaled Tanh activation function.  The Linear Channel and Scaled Tanh are defined in Section~\ref{sec:wgan-gp}. Note that a transposed convolution upsamples an image by a factor of two. 
  }
  \label{fig:wgan-gp-generator_summary}
  \vspace{2mm}
\end{figure*}

The DCWGAN-GP model we use is summarized by the generator depicted in Figure~\ref{fig:wgan-gp-generator_summary} and the discriminator described in Table~\ref{tab:wgangp-discriminator_summary}.  We take the {\it CosmoGAN} network architecture used in~\cite{Mustafa2019} as a starting point of our DCWGAN-GP network because it has been demonstrated to reproduce non-Gaussian statistics for a one-component map.  From this starting point, we make the following adjustments. First, we switch the simple GAN loss function used in {\it CosmoGAN} with the Wasserstein loss function with gradient penalty introduced in~\cite{Gulrajani2017}; the loss function is the statistic that all GANs optimize.  We find that this change of the loss-function substantially improves the stability of our network against mode-collapsing (i.e.~against the network missing critical features). To be consistent with the standard Wasserstein GAN (WGAN) architecture, we remove the batch normalization layers (i.e.~the intermediate normalization steps within the GAN) from the discriminator to stabilize the loss function gradient, and replace the sigmoid function activation layer with a linear function activation layer, following~\cite{Gulrajani2017}. In addition, we increase the depth of the convolution layers in both generators and discriminators from four to five, and we increase the latent-vector length (i.e.~array size of input random numbers) from the original 64 to 256 in order to have a more expressive network.  We also change the last activation layer in the generator model from $\tanh{x}$ to $a*\tanh{b/a*x}$, which we call a Scaled Tanh. Here $a$ is chosen to be 15 to accommodate the dynamic range of the normalized pixel values we obtained using Equation~\ref{eqn:scaled_log_norm}, while $b$ is chosen to be $2$ so that the gradient around the boundary values (i.e. -15 and 15) do not vanish too rapidly; we find the form of the Scaled Tanh function and its parameters by checking the training statistics when using the validation set of patches (see Section~\ref{sec:prep2}). We also add a custom linear layer that linearly re-scales each extragalactic component independently so that the network has more freedom to re-scale the components relative to each other. We call this custom layer the "Linear Channel" (LC). We find that using this LC helps the network to converge faster.  We reduce the two-dimensional convolution layer kernel size (i.e.~the width and the height of a convolution filter) from $5\times 5$, used in {\it CosmoGAN}, to $4\times 4$ to account for the difference in the input image size from $256\times 256$ pixels to $128\times 128$ pixels (the latter is for $1^{\circ}$ x $1^{\circ}$ patches at 0.5 arcmin resolution). Lastly, we replace the Rectified Linear Unit (ReLU) layers in the generator with the Leaky Rectified Linear Unit (Leaky ReLU) with $\alpha=0.2$ to achieve better performance, as suggested by~\cite{Xu2015}. (For more discussion about various types of activation functions, we refer to~\cite{Nwankpa2018}). 

We train the DCWGAN-GP network with the Primary Input Data using up to 100 epochs (i.e.~the number of cycles through the entire training data set used by the network during training).  Following~\cite{Gulrajani2017}, we adopt a gradient penalty coefficient ($\lambda$) of 10 and the number of discriminator iterations per generator iteration ($n_{critic}$) of 5. For gradient decent, we use the Adam optimizer~\cite{Kingma2014} with a learning rate ($lr$) of $10^{-4}$ and $(\beta_1, \beta_2)=(0.5,0.9)$. Throughout this training, we check not only the loss function values, but also the visual images, Minkowski functionals, power spectra, cross spectra, and pixel PDFs of the network outputs; we will discuss each metric in Section~\ref{sec:results}. If the network shows signs of over-fitting or mode-collapsing, we restart the training either from the beginning or after reducing the learning rate by half (i.e.~effectively reducing the step size). We terminate the training cycle once all the statistics considered are satisfactorily reproduced. We use the trained DCWGAN-GP network to produce {\it Intermediate Product 1} (800,000 $1^{\circ}$ x $1^{\circ}$ patches of non-Gaussian $\kappa$ maps and correlated non-Gaussian foregrounds) which we use in Training Step 2. Note that the samples in Intermediate Product 1 are no longer overlapping and are statistically independent of each other. 
 
\begin{table}[t]
\begin{tabular}{|c|c|c|c|}
\hline
Layer            & Activation      & Output Shape & $\#$ of Trainable \\
                 & Function      & of Tensor & Parameters \\
\hline
Input Map        & N/A    & $N$x128x128    &  N/A      \\ \hline
Conv 4x4         & LReLU  & 64x64x64     & 5,184       \\ \hline
Conv 4x4         & LReLU  & 128x32x32    & 131,200      \\ \hline
Conv 4x4         & LReLU  & 256x16x16    & 524,544      \\ \hline
Conv 4x4         & LReLU  & 512x8x8      & 2,097,664       \\ \hline
Conv 4x4         & LReLU  & 1024x4x4     & 8,389,632       \\ \hline
Linear           & N/A      & 1            & 16,385       \\ \hline
\multicolumn{3}{|c|}{Total Trainable Parameters}  & 11,164,609       \\ \hline
\end{tabular}\caption{Summary of the discriminator model described in Section~\ref{sec:wgan-gp}. We use the same discriminator model throughout all three training steps described in Section~\ref{sec:method}. The first and second columns list the layers and the corresponding activation functions. The third and the fourth columns give the shapes of the output arrays and the number of trainable parameters in each layer. Here, $N$ is the number of input components, i.e.~$N=5$ for DCWGAN-GP and PIXGAN and $N=10$ for VAEGAN. Conv $4\times 4$ indicates a two-dimensional convolution layer with a $4\times 4$ kernel, and LReLU stands for a leaky rectified linear unit activation with $\alpha=0.2$ (see Section~\ref{sec:wgan-gp} for details). Note that Conv $4\times 4$ downsamples an image by a factor of two, and increases the number of features to consider.}
\label{tab:wgangp-discriminator_summary}
\end{table}

\subsection{Training Step 2: Generate Extragalactic Foregrounds from Kappa Map with Pix2Pix GAN (PIXGAN)}
\label{sec:pix2pix}
\begin{figure*}[t]
  \centering
  \hspace{-3mm}\includegraphics[width=\textwidth]{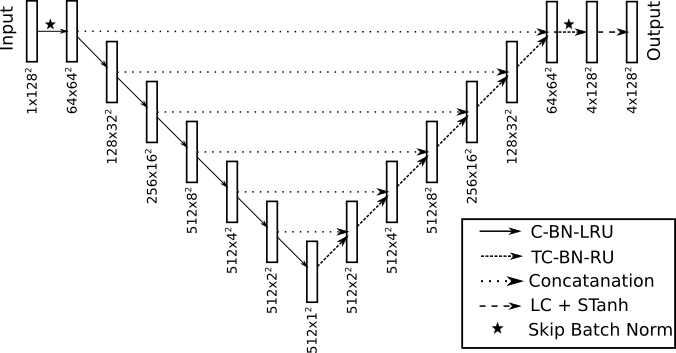}
  \caption{Shown is a diagram of the UNET generator model described in Section~\ref{sec:pix2pix} and indicated in the schematic of Figure~\ref{fig:flowchart} as Training Step 2. The generator takes as input a $1^{\circ}$ x $1^{\circ}$ $\kappa$ map, and it outputs a set of $1^{\circ}$ x $1^{\circ}$ extragalactic foreground maps consisting of tSZ, kSZ, CIB, and Radio components. The direction of the data flow is indicated by the arrows. A solid arrow, labeled C-BN-LRU, represents a sequence of 1)~a $4\times 4$ two-dimensional convolution, 2)~a batch normalization, and 3)~a leaky rectified linear unit activation function. A short-dashed arrow, labeled TC-BN-RU, represents a sequence of 1)~a $4\times 4$ two-dimensional transposed convolution, 2)~a batch normalization, and 3)~a rectified linear unit activation function. A two-dimensional convolution downgrades an image by a factor of two, while a transposed convolution upsamples an image by the same factor. A dotted arrow is a concatenation operation, which concatenates the output of an originating box to the output of a destination box. A long-dashed arrow, labeled LC + STanh, on the top right represents a sequence of a Linear Channel and a Scaled Tanh activation function. Linear Channel and Scaled Tanh are defined in Section~\ref{sec:wgan-gp}. Deviating from our notation, we do not apply a batch normalization to the first and the last convolution layers, indicated by a star over a line.
  }
  \label{fig:pix2pix-generator_summary}
  \vspace{2mm}
\end{figure*}

The PIXGAN model we use is summarized by the generator depicted in Figure \ref{fig:pix2pix-generator_summary} and the discriminator described in Table \ref{tab:wgangp-discriminator_summary}. A PIXGAN can convert input images to other images by implementing a U-NET  generator~\cite{Ronneberger2015, Isola2016}. We use the PIXGAN network to convert a given $\kappa$ map to the other four non-Gaussian extragalactic foregrounds consisting of tSZ, kSZ, CIB, and Radio components. We start with the original PIXGAN architecture presented in~\cite{Isola2016}, and make the following modifications.
We first replace the $\tanh{}$ activation layer in the generator with a sequence of a Linear Channel and a Scaled Tanh; the latter two are defined in \ref{sec:wgan-gp}.  Since we terminate the training well before the network shows signs of over-fitting, we also remove the drop-out layers from the generator to speed up the training. Lastly, instead of using the PatchGAN discriminator as in~\cite{Isola2016}, we use the same DCWGAN-GP discriminator model described in Section \ref{sec:wgan-gp}, and summarized in Table~\ref{tab:wgangp-discriminator_summary}.

We use Intermediate Product 1 (800,000 $1^{\circ}$ x $1^{\circ}$ non-overlapping patches of non-Gaussian $\kappa$, tSZ, kSZ, CIB and Radio maps) generated by Training Step 1 to train the PIXGAN network.{\color{blue}\footnote{We still use the test and the validation data sets from our Primary Input Data to fine-tune the network parameters and to validate the network outputs.}} We use the same optimizer setup as in Training Step 1. During the training process, we find that the PIXGAN network tends to overestimate the number of massive tSZ clusters, which results in an excess of tSZ power at large scales. To mitigate this effect, we add a linear loss term to the DCWGAN-GP generator loss function evaluation defined as the second term in the right side of the equation below. 
 
\begin{equation}
    \begin{aligned}
     \mathcal{L}^{PIXGAN}_G &= \mathcal{L}^{DCWGAN-GP}_G \\ 
     &+ \gamma \sum_i|\sigma^2(S^i_{\rm{network}})-\sigma^2(S^i_{\rm{input}})| 
     \end{aligned}
     \label{eqn:linearLoss}
\end{equation}
where $\mathcal{L}^{DCWGAN-GP}_G$ is the standard DCWGAN-GP generator loss function defined in~\cite{Gulrajani2017}, and $\sigma^2(S^i)$ is the total pixel variance of the i-th foreground component in a sample.  Since the sum of power spectra values across multipoles is the equal to the total pixel variance of the map (a corollary of Parseval's theorem), it follows that penalizing the difference in the total pixel variance is equivalent to penalizing the mismatch in the summed power spectra. We train the model with $\gamma =100$, and find that it helps to regulate the excess power in the tSZ map. 
 
We train the PIXGAN network using up to 10 epochs~{\color{blue}\footnote{Note that each epoch here has fours times more samples than the epochs in Section~\ref{sec:wgan-gp}}}. We validate the network every 200,000 training samples (four times per epoch) with the updated loss function in Equation~\ref{eqn:linearLoss}, checking the same metrics as in Section~\ref{sec:wgan-gp}. We intentionally terminate early the training of the PIXGAN network for two reasons. First, once the PIXGAN network starts to overfit, the quality of the output degrades quickly. We find it more conservative to under-train the PIXGAN network rather than to risk over-fitting. Second, we find that the VAEGAN network described in the next training step (Training Step 3) can absorb some imperfection in the simulations and correct for them. Therefore, the output from the PIXGAN network can have minor defects without affecting the quality of the final products. We terminate the training once we see signs of over-fitting; these signs are 1)~minimal improvement of the loss function over the training samples, 2)~large fluctuations of the other metrics with respect to the S10 simulations, and 3)~images missing significant features.  At this point, we choose the snapshot of the network that reproduces the summary statistics the best.  In particular, for the power spectra, we check that the PIXGAN network can reproduce the shape of each foreground component power spectra up to a constant multiplication factor.{\color{blue}\footnote{Note that the next network (Training Step 3) can re-scale each foreground component by a constant factor as described in Section~\ref{sec:vaegan}.}} When this network snapshot reproduces the shape of the power spectra to within $10\%$ for $\ell \in (2000,8000)$, and the cross spectra and Minkowski functionals also show similar agreement, we consider the training complete. Otherwise, we repeat the training procedure until we reach this criteria.
After this training step is done, we feed in several full-sky Gaussian $\kappa$ maps through the trained PIXGAN network, producing 200,000 $1^{\circ}$ x $1^{\circ}$ non-overlapping patches of extragalactic foregrounds consisting of tSZ, kSZ, CIB, and Radio components; we call this {\it Intermediate Product 2}. The full-sky Gaussian $\kappa$ maps are generated using {\it pixell}; we use the power spectra of the S10 kappa map as the input spectra to these Gaussian kappa simulations.

\begin{figure*}[t]
  \centering
  \hspace{-3mm}\includegraphics[width=\textwidth]{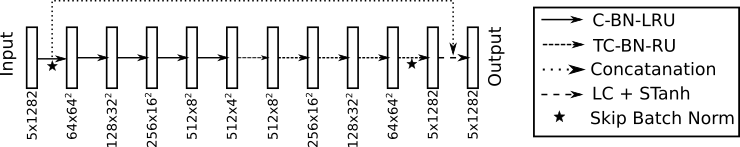}
  \caption{Shown is a diagram of the VAEGAN generator model described in Section~\ref{sec:vaegan} and indicated in the schematic of Figure~\ref{fig:flowchart} as Training Step 3. The input to this generator is a set of $1^{\circ}$ x $1^{\circ}$ {\it{Gaussian}} $\kappa$ maps and their corresponding extragalactic foreground maps from the PIXGAN network in Figure~\ref{fig:pix2pix-generator_summary} (i.e.~Intermediate Product 2). The output are a set of non-Gaussian $\kappa$ maps and their corresponding non-Gaussian extragalactic foreground maps. The direction of the data flow is indicated by arrows. A solid arrow, labeled C-BN-LRU, represents a sequence of 1)~a $4 \times 4$ two-dimensional convolution, 2)~a batch normalization, and 3)~a leaky rectified linear unit activation function. A short-dashed arrow, labeled TC-BN-RU, represents a sequence of 1)~a $4\times 4$ two-dimensional transposed convolution, 2)~a batch normalization, and 3)~a rectified linear unit activation function. The two-dimensional convolution downgrades the image by a factor of two, while the transposed convolution upsamples the image by the same factor.  Shown as a dotted arrow is a concatenation operation between the input Intermediate Product 2 and the VAEGAN network output, which focuses the network on the residual difference between the two; note that we apply the inverse of Scaled Tanh to this input prior to the concatenation.  A long-dashed arrow, labeled LC + STanh, represents a sequence of a Linear Channel and a Scaled Tanh activation function; they are described in Section~\ref{sec:wgan-gp}.  Deviating from our notation, we do not apply a batch normalization to the first and the last convolutions layers, indicated by a star below the arrow.
  }
  \label{fig:vaegan-generator_summary}
  \vspace{2mm}
\end{figure*}

\subsection{Training Step 3: Restoration of Non-Gaussian Information
with Variational Autoencoder GAN (VAEGAN)}
\label{sec:vaegan}

The VAEGAN model we use is summarized by the generator depicted in Figure~\ref{fig:vaegan-generator_summary} and the discriminator described in Table~\ref{tab:wgangp-discriminator_summary}.  Neural networks can be used to improve existing simulations, either by improving the resolution or by adding missing small-scale information~\cite{Dong2014, Krachmalnicoff2020}. In this work, we adopt the Variational Autoencoder GAN (VAEGAN) in order to add back missing non-Gaussian information~\cite{Larsen2015}. Our VAEGAN generator architecture is similar to the U-NET generator architecture discussed in Section~\ref{sec:pix2pix} except for the following. Unlike the PIXGAN, the VAEGAN generator does not have skip-connections for each layer (i.e.~there are no array concatenations for each layer like as shown by the dotted lines in Figure~\ref{fig:pix2pix-generator_summary}), except that we add back input maps from Intermediate Product 2 at the very end (see dotted line in Figure~\ref{fig:vaegan-generator_summary}) as done in~\cite{Thiele2020}. This addition at the end allows the VAEGAN network to focus just on the missing non-Gaussian information (i.e.~to focus on the residual difference between the network product of Training Step 3 and Intermediate Product 2). Another difference between the PIXGAN and VAEGAN generators are that the latter has fewer trainable parameters, which makes it easier to train; thus we only need 200,000 samples, as opposed to 800,000, to train it. Regarding the discriminator, we use the same DCWGAN-GP discriminator (i.e.~the same Wasserstein loss function 
with the gradient penalty) discussed in Section~\ref{sec:wgan-gp}, and summarized in Table~\ref{tab:wgangp-discriminator_summary}.

We use as input both the Primary Input Data and Intermediate Product 2 to train the VAEGAN network. The output of the VAEGAN network is the restoration of the missing non-Gaussian information in Intermediate Product 2.  The VAEGAN network does this by comparing the original Intermediate Product 2 to the S10 simulations and learning the difference.  To do this we use the same loss function and training setup as in Training Step 2.  Since the output from Training Step 3 is our final product, except for a few corrections described later in Section~\ref{sec:postprocess}, we further fine-tune the Training Step 3 training procedure to achieve optimal results. In particular, we revisit the issue of overestimation of massive tSZ clusters that we described in Section~\ref{sec:pix2pix}. In order to further reduce the chance of overestimation, we make the following adjustments to our training procedure: 1)~first, we sort the training samples by the total variance of the tSZ component, 2)~then we start training with the bottom eighty percent of samples in terms of the total tSZ variance; 3)~Once the training is stabilized, we restart the training with the full set of training samples.  We closely monitor the tSZ statistics along with the loss function and other summary statistics. We terminate the training when all the statistics considered are within $5\%$ for $\ell \in (2000, 8000)$.

\subsection{Post-processing the Network Simulations}
\label{sec:postprocess}

As discussed in Section~\ref{sec:method}, we convert a full-sky Gaussian $\kappa$ map to other foreground components tile by tile, where each tile is a $1^{\circ}$ x $1^{\circ}$ flat patch of sky. Here, we describe our method to divide full-sky $\kappa$ maps into tiles and to reproject the processed tiles back on to a full-sky map. When we sample a $1^{\circ}$~x~$1^{\circ}$ patch from a full-sky $\kappa$ map, we rotate that patch to the celestial equator and then cut out a tile of size $128 \times 128$ pixels with a 20-pixel overlap region between each tile as shown in Figure~\ref{fig:tile_layout} in Appendix~\ref{sec:grid}. This is similar to how we cut the original S10 simulations as discussed in Section~\ref{sec:prep2}. This results in a grid of tiles with 200 tiles along a longitudinal direction and 400 tiles along a latitude direction.{\color{blue}\footnote{Around the poles of the sphere (i.e.~the first and the last longitudinal rows of the grid), there is not enough physical area to sample a $1^{\circ}$~x~$1^{\circ}$ tile. Hence, we take the second and the second-to-the-last rows of the grid, mirror them longitudinally, and replace the first and the last rows of the grid with these mirrored rows. This ensures that we have continuous physical tiles around the poles at the cost of them not being unique. We find that this pole correction does not affect the overall statistics of the network simulations.}} When we reproject the tiles back on to a full-sky map, we interpolate them back to the correct location on the sphere according to their coordinates. Here, we take advantage of the fact that CAR pixels in a full-sky map get stretched only along the latitude direction. As a result, the pixels in the longitudinal direction can be remapped without any interpolation. This effectively reduces the dimensions of the interpolation from two to one, which lowers the overall computational cost. 

We use two different interpolation methods depending on which foreground component we would like to reproject.  We use a linear interpolation to reproject the $\kappa$, tSZ, and kSZ foreground components because they are continuous fields.{\color{blue}\footnote{We tried both linear interpolation and popular cubic spline methods, and found no noticeable difference in the resulting summary statistics. Thus, we chose to use the linear interpolation method which is more computationally efficient.}} To reproject the CIB and Radio galaxies, we use the nearest neighbor method (i.e.~each CAR pixel in the full-sky map obtains a value from the nearest pixel in the flat-sky tile).  We use this method because linear interpolation can "blur" sharp variations in amplitude, which is not ideal for interpolating discrete point sources. We remove any superfluous negative flux sources before reprojecting point source tiles back onto the full-sky. Since we initially sample the full-sky $\kappa$ map with overlapping regions, one pixel in the full-sky map can be mapped onto more than one flat-sky tile. Therefore, we precompute the number of mapped tiles for each full-sky map pixel and construct an effective ``hit-counts'' map. If more than one tile corresponds to a single full-sky pixel, then we add the tile pixel values and divide by the hit-counts map to obtain an average.  In addition, we apodize each tile with a cosine apodization mask prior to the full-sky reprojection as done in~\cite{Krachmalnicoff2020} to suppress any discontinuity around the edges; this is done by weighting the overlapping regions by the apodization mask to downweight pixels closer to the edges of a given tile.{\color{blue}\footnote{We set the pixel values at the outermost edges of the apodization mask to the values of pixels one pixel away inside the map to avoid losing information.}} Lastly, following Section 4.3 in~\cite{Naess2020}, we compute two-dimensional transfer functions to correct for the reprojection effects for the $\kappa$, tSZ, and kSZ maps. Note that we do not apply two-dimensional transfer functions to the CIB and Radio maps to avoid spectral leakage when shifting to Fourier space.

We find that the raw full-sky outputs from the network can reproduce the statistics of the S10 simulations overall quite well. However, we do find some small mismatches in the summary statistics between the network and the S10 simulations; for example, for the power spectra in the range $\ell \in (2000, 8000)$, the agreement is better than $3\%$, $10\%$, $0.5\%$, $17\%$ and $0.3\%$ for $\kappa$, kSZ, tSZ, CIB and Radio respectively, however, it is not perfect.  Therefore, we make a few corrections to the network simulations to match the S10 simulations more closely. Here, we describe the corrections we apply. First, to correct for the small mismatch in the power spectra, we apply one-dimensional transfer functions to the $\kappa$, kSZ, and tSZ maps. These transfer functions are computed as the square root of the ratio of the average of 10 network power spectra to the S10 power spectra. We compute the transfer functions at each multipole, and apply to these transfer functions a one-dimensional Gaussian filter with $\sigma_{\ell}=10$ to smooth them out. We then convert these to isotropic two-dimensional transfer functions, and apply them to the spherical harmonics of the full-sky map.

We find larger deviations in the power spectrum between the network and the S10 simulations than specified above at larger ($\ell \le 2000$) and smaller scales ($\ell \ge 8000$). The large scale deviation is due to the limited size of the sample patches as discussed in Section~\ref{sec:method}. For the small scales, we find a roll-off in the power spectra starting at $\ell = 8000$. We find that this roll-off starts at larger scales ($\ell \le 8000$) if we train the network with lower resolution simulations (for example, 1 arcminute resolution), which indicates that the roll-off is related to the resolution of the initial training samples. 

A second correction we apply is to adjust the CIB and Radio maps as follows. We re-scale the CIB map by a factor of 1.1 to match the S10 power spectra for $\ell \in (4000, 8000)$. We also re-scale the network CIB and Radio source fluxes to match the re-scaled S10 source counts.{\color{blue}\footnote{Note that the original S10 CIB source fluxes have been re-scaled by a factor of 0.75 as discussed in Section~\ref{sec:prep1}.}} After this re-scaling, we still find that the network overestimates the number counts of CIB sources with flux values greater than 1~mJy; thus we reduce the flux of those sources by replacing the original flux value, $S$, with $S^{0.63}$. For Radio sources, we find that replacing the original flux by $S^{(\lambda)} = ((S+1)^{\lambda}-1)/\lambda$ with $\lambda = 1.25$~\cite{Yeo2000} improves the match at the high-flux end.

\subsection{Generating Lensed CMB and Multi-frequency Maps}
\label{sec:}

In order to generate lensed CMB ($T,Q,U$) maps, we first generate full sky unlensed CMB ($T,Q,U$) maps as Gaussian Random Fields on a CAR full-sky grid using the CAMB theory spectra used in S10.{\color{blue}\footnote{The CAMB theory spectra used in this work can be found under the resources directory at https://github.com/dwhan89/cosmikyu.}} Then, these unlensed CMB maps are lensed with a network full-sky $\kappa$ map using the lensing method described in~\cite{Choi2020}. As in~\cite{Choi2020}, the lensing operation is performed at 1 arcminute resolution and agrees with the theory to better than few percent up to $\ell = 10,000$. 

Up to this point, all the training and corrections are done with the 148~GHz network simulations. Here, we discuss how we simulate the other frequency maps.  We assume to leading order that the lensed CMB and kSZ perfectly follow a black-body spectrum (i.e.~they do not vary with frequency in units of differential CMB temperature). The $\kappa$ map is frequency independent. In order to simulate the tSZ map at different frequencies, we use the analytic calculations in~\cite{Zeldovich1969, Sunyaev1970} to compute the frequency dependent scaling factor, $a(\nu)$, and scale the 148~GHz tSZ map following $M^{{\rm{tSZ}}}({\nu}) = a(\nu)/a(148)*M^{\rm{tSZ}}({148~GHz})$, where $M^{\rm{tSZ}}$ is the real-space tSZ map. For CIB and Radio sources, we find the mean spectral indices between each frequency and the reference frequency of 148~GHz, directly from the S10 catalogues; this results in the spectral indices shown in Table~\ref{tab:freq} of Appendix~\ref{sec:freq}. For each source in the 148~GHz map, we then randomly sample a spectral index from a Gaussian distribution with the mean and the standard deviation given in Table~\ref{tab:freq}, and scale its flux accordingly. In this way, we generate full-sky maps of the combined lensed CMB, kSZ, tSZ, CIB, and Radio signals at 30, 90, 148, 219, 277, and 350~GHz.

\begin{figure*}[t]
  \centering
  \hspace{-3mm}\includegraphics[width=\textwidth,height=\textheight,keepaspectratio]{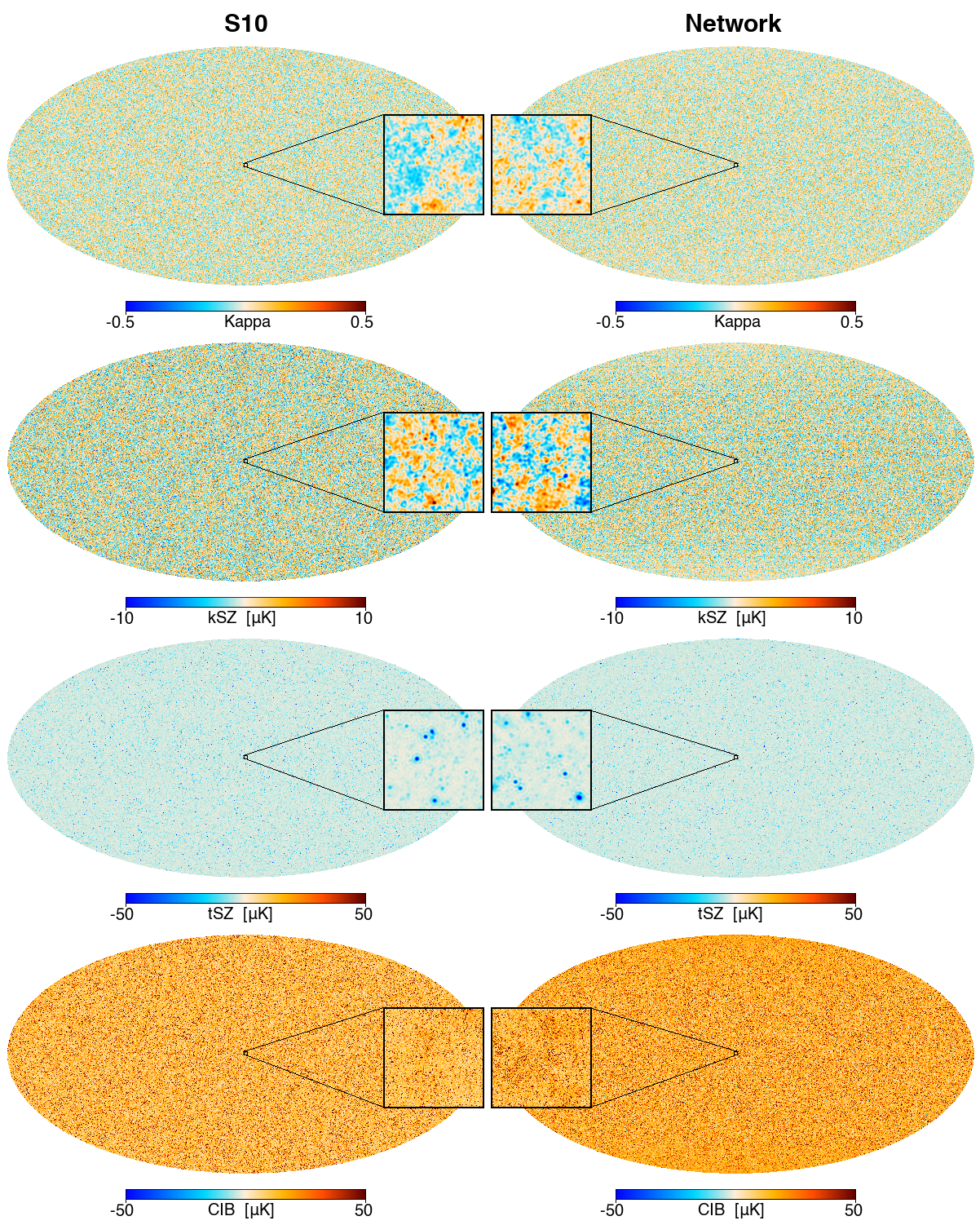}
  \caption{From top to bottom, shown are the lensing convergence ($\kappa$), the kinetic Sunyaev-Zel'dovich effect (kSZ), the thermal Sunyaev-Zel'dovich effect (tSZ), and the Cosmic Infrared Background (CIB) maps at 148~GHz from the S10 simulations (left column) and from the network (right column). A flux cut of 7~mJy at 148~GHz is applied to the CIB maps. Full-sky maps are shown in the Mollweide projection in the background, while center panels show zoom-ins of $1^{\circ}$ x $1^{\circ}$ patches.  All maps have the units of $\mu$K, except for the $\kappa$ map which is dimensionless. Note that the S10 simulations are unique for only one octant of the sky; on the other hand, the network simulations do not have any repeated tiles.}
  \label{fig:maps}
\end{figure*}

\section{Results}
\label{sec:results}

In this section, we study the statistical properties of the mmDL simulations generated by our network. Unless otherwise noted, all statistics are computed using full-sky maps after applying a flux-cut of 7~mJy at 148~GHz to Radio and CIB maps, and appling the corrections discussed in Section~\ref{sec:postprocess}. We first visually compare the full-sky mmDL and S10 simulations in Figure~\ref{fig:maps}. Here we show the full-sky maps in the Mollweide projection, as well as zoom-ins of $1^{\circ}$ x $1^{\circ}$ patches. Visually, the mmDL simulations and the S10 simulations are indistinguishable. 

\begin{figure}[t]
  \centering
  \hspace{-3mm}\includegraphics[width=0.91\columnwidth]{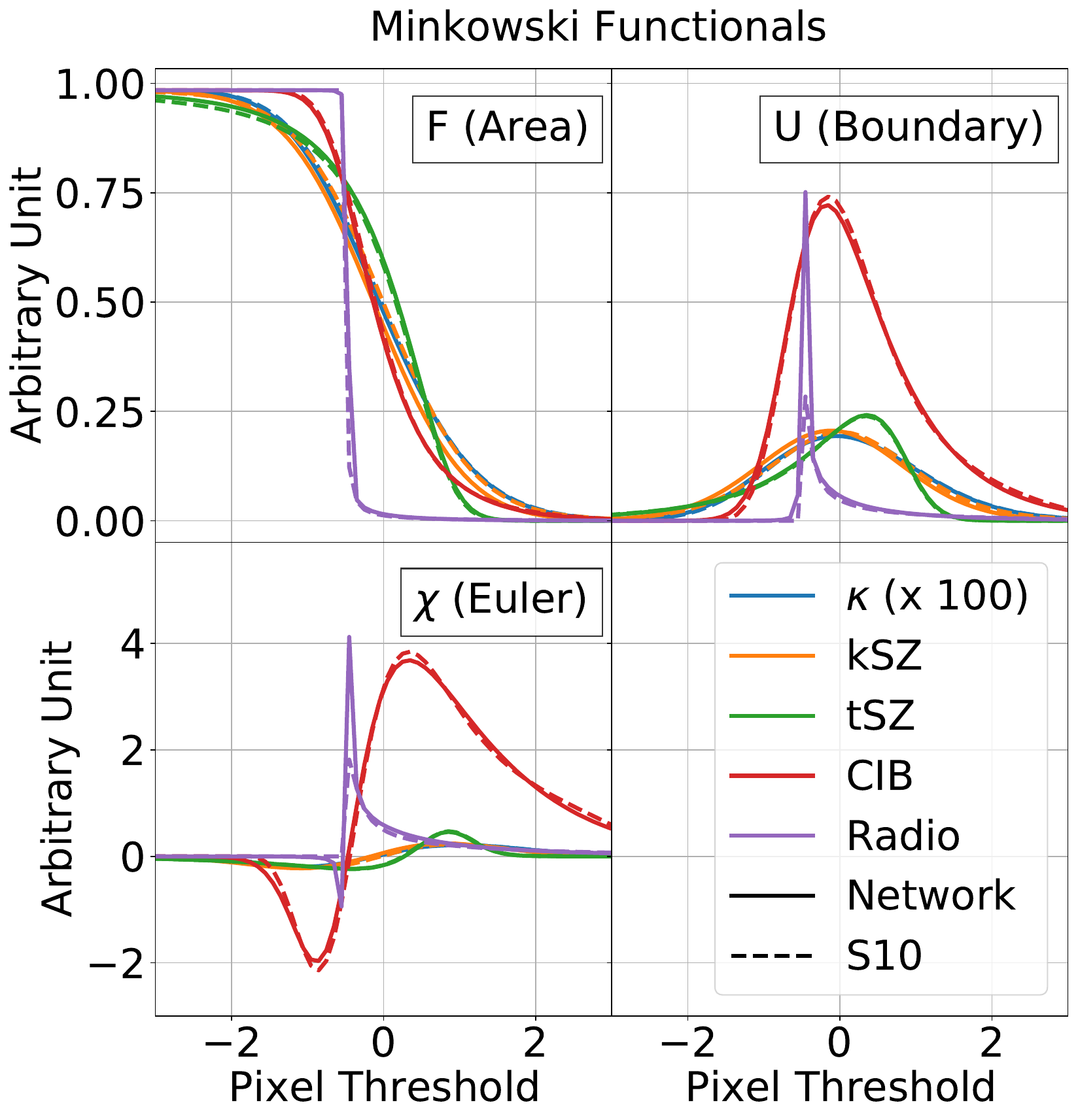}
  \caption{Shown are the Minkowski functionals of the extragalactic foreground components at 148~GHz. The solid curves are calculated using the average of 100 $1^{\circ}$ x $1^{\circ}$ tiles output by the VAEGAN described in Section~\ref{sec:vaegan}. The dashed curves are calculated using the full-sky S10 simulation described in Section~\ref{sec:dataset}. Both the S10 and the network simulations are normalized following the procedure detailed in Section~\ref{sec:prep3}.  To compute the Minkowski functionals, a binary map is first constructed with a value of one above a certain pixel threshold, and zero otherwise. Then, the this binary map is used to compute the area above the threshold ($F$), its boundary ($U$), and the dimensionless curvature known as the Euler characteristic ($\chi$). These three functionals, give an estimate of the non-Gaussianity of each component~\cite{Mantz2008}. We find that the network can reproduce well the Minkowski functionals calculated using the S10 simulations. This indicates that the network learned the non-Gaussian information in S10.}
  \label{fig:minko}
\end{figure}

\subsection{Pixel-level Summary Statistics}

\begin{figure}[t]
  \centering
  \hspace{-3mm}\includegraphics[width=1\columnwidth]{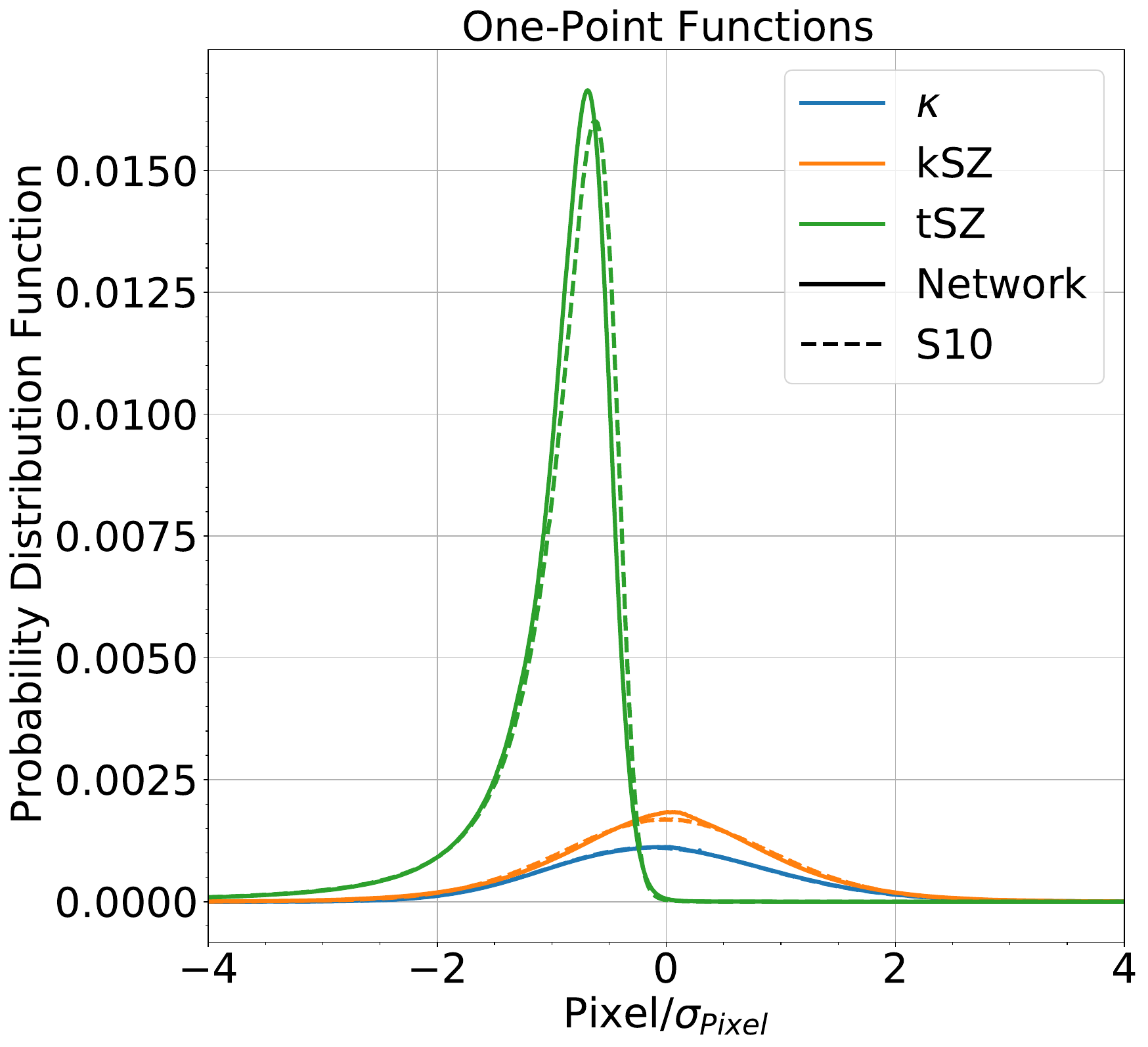}
  \caption{Shown are one-point probability distribution functions of the extragalatic foregrounds at 148~GHz. Solid curves are calculated from a random sample of a full-sky simulation from the network, and dashed curves are calculated using the full-sky S10 simulation. A flux-cut of 7~mJy is applied to both CIB maps. Each map is divided by the standard deviation of its pixel values in order to have unit variance.  We find good agreement between the network curves and the S10 curves. 
  }
  \label{fig:onepoint_all}
\end{figure}

There are a number of pixel-level summary statistics used in the literature.  Such examples are 1)~Minkowski functionals~\cite{Mantz2008}, 2)~one-point probability distributions, and 3)~source counts. For two dimensional images, Minkowski functionals measure the area ($F$), the contour length ($U$), and the dimensionless curvature known as the Euler characteristic ($\chi$) as a function of varying pixel threshold value.  Combined together these three functionals give an estimate of the topology of an image, which is sensitive to the presence of non-Gaussianity. We compute Minkowski functionals using the {\it minkfncts2d} software, which is publicly available at \href{https://github.com/moutazhaq/minkfncts2d}{https://github.com/moutazhaq/minkfncts2d}. Figure~\ref{fig:minko} shows as solid curves the functionals computed using the average of 100 $1^{\circ}$ x $1^{\circ}$ tiles output by the VAEGAN described in Section~\ref{sec:vaegan}.{\color{blue}\footnote{We recomputed the Minkowski functionals on a 20 degree wide band around the celestial equator after making the corrections described in Section~\ref{sec:postprocess}; we did not find any significant change, except a slight shift of the CIB curves as compared to Figure~\ref{fig:minko}.}} The dashed curves are calculated using the full-sky S10 simulation described in Section~\ref{sec:dataset}.  Here, we compute the functionals from the network without the corrections discussed in Section~\ref{sec:postprocess}. Hence, the Minkowski functionals shown in Figure~\ref{fig:minko} give an estimate of how well the network performs before applying any corrections. Overall, we find very good agreement between the functionals computed from the network and the S10 simulations. The only noticeable mismatch is in the $U$ and $\chi$ functionals of the Radio sources, which indicates a mismatch in the real-space shape of Radio point sources between the mmDL and S10 simulations. The S10 point sources are "true" point sources by construction (i.e. each source falls into a single pixel). On the other hand, the network point sources are constructed as the summation of multiple convolution kernels, hence they can potentially extend beyond a single pixel. In practice, the simulations will be convolved with a beam with a full width at half maximum (FWHM) larger than 0.5 arcminute. So this minor mismatch in the Radio source shape will have a negligible effect in actual analyses. 

\begin{figure}[t]
  \centering
  \hspace{-3mm}\includegraphics[width=0.95\columnwidth]{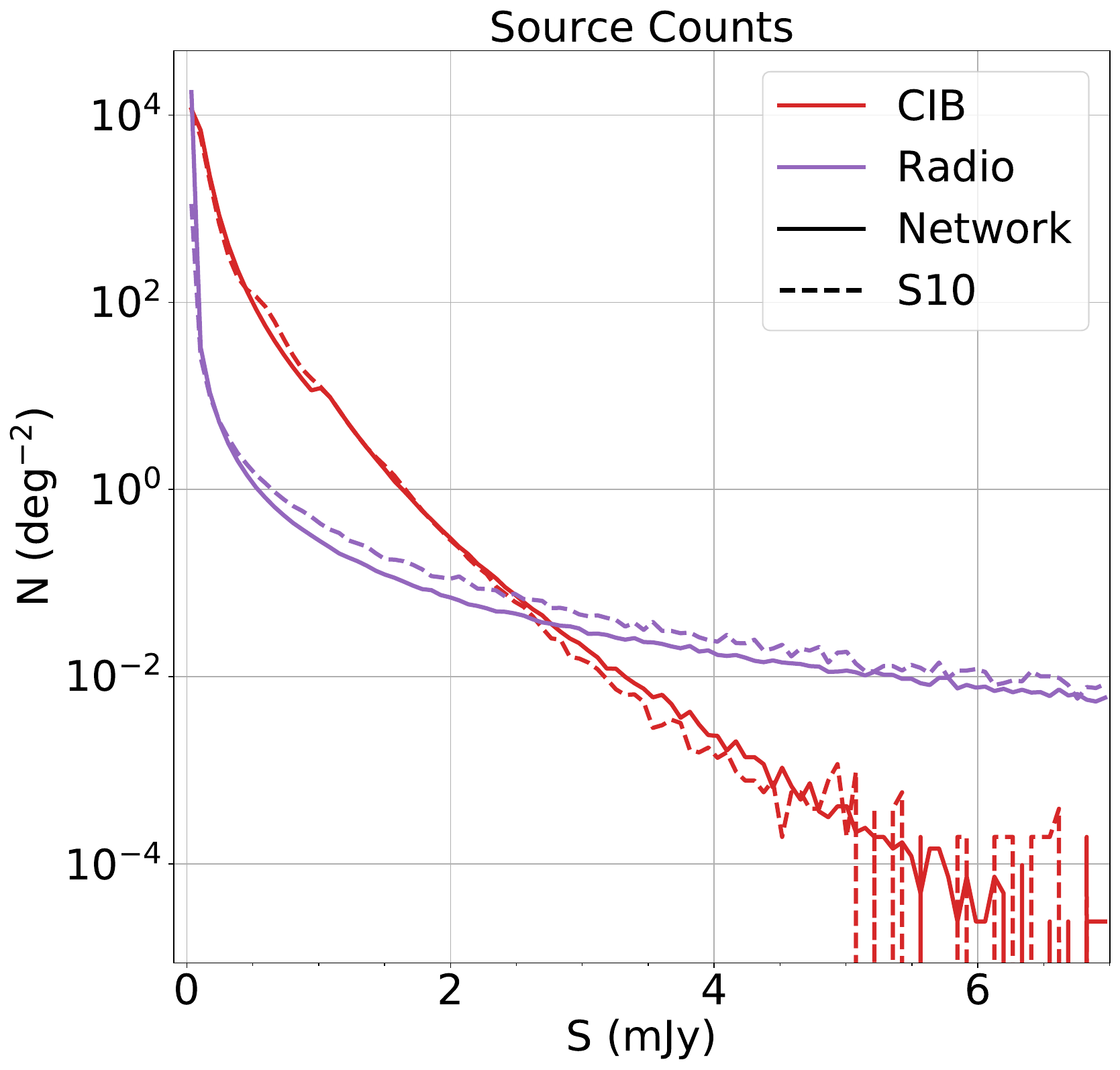}
  \caption{Shown is a histogram of the source number counts per square degree per flux bin for CIB and Radio maps. The solid curves are generated using a random simulation from the network, while the dashed curves are generated using the S10 simulation. A flux cut of 7~mJy is applied to both CIB and Radio maps for both network and S10 simulations.  We find good overall agreement. } 
  \label{fig:source}
\end{figure}

Another pixel-level summary statistic we consider is the one-point probability distribution function (one-point PDF) shown in Figure~\ref{fig:onepoint_all}; this one-point PDF is the histogram of pixel values. Before computing this histogram, we divide the pixels in each component map by the standard deviation of pixel values in that map in order to have unit variance. One-point PDFs have been studied previously in the context of CMB weak lensing and tSZ analyses (e.g.~\cite{Hill2014, Liu2016, Thiele2020}), and have been shown to be good probes of non-Gaussian information. We find good overall agreement between a random realization of a full-sky network simulation and the S10 simulations.

The last pixel-level statistics we consider is the source number counts. In Figure~\ref{fig:source}, we show the histogram of the source number counts per square degree per flux bin for the simulated CIB and Radio maps. As discussed in Section~\ref{sec:postprocess}, the network overestimates the number of CIB sources with flux levels greater than 1~mJy, and we correct for this by scaling down the network flux levels at the high flux tail. This scaling substantially improves the overall match. This scaling has negligible impact on the other CIB statistics discussed below, since the other statistics are largely driven by lower flux sources.

\begin{figure}[t]
  \centering
  \hspace{-3mm}\includegraphics[width=0.95\columnwidth]{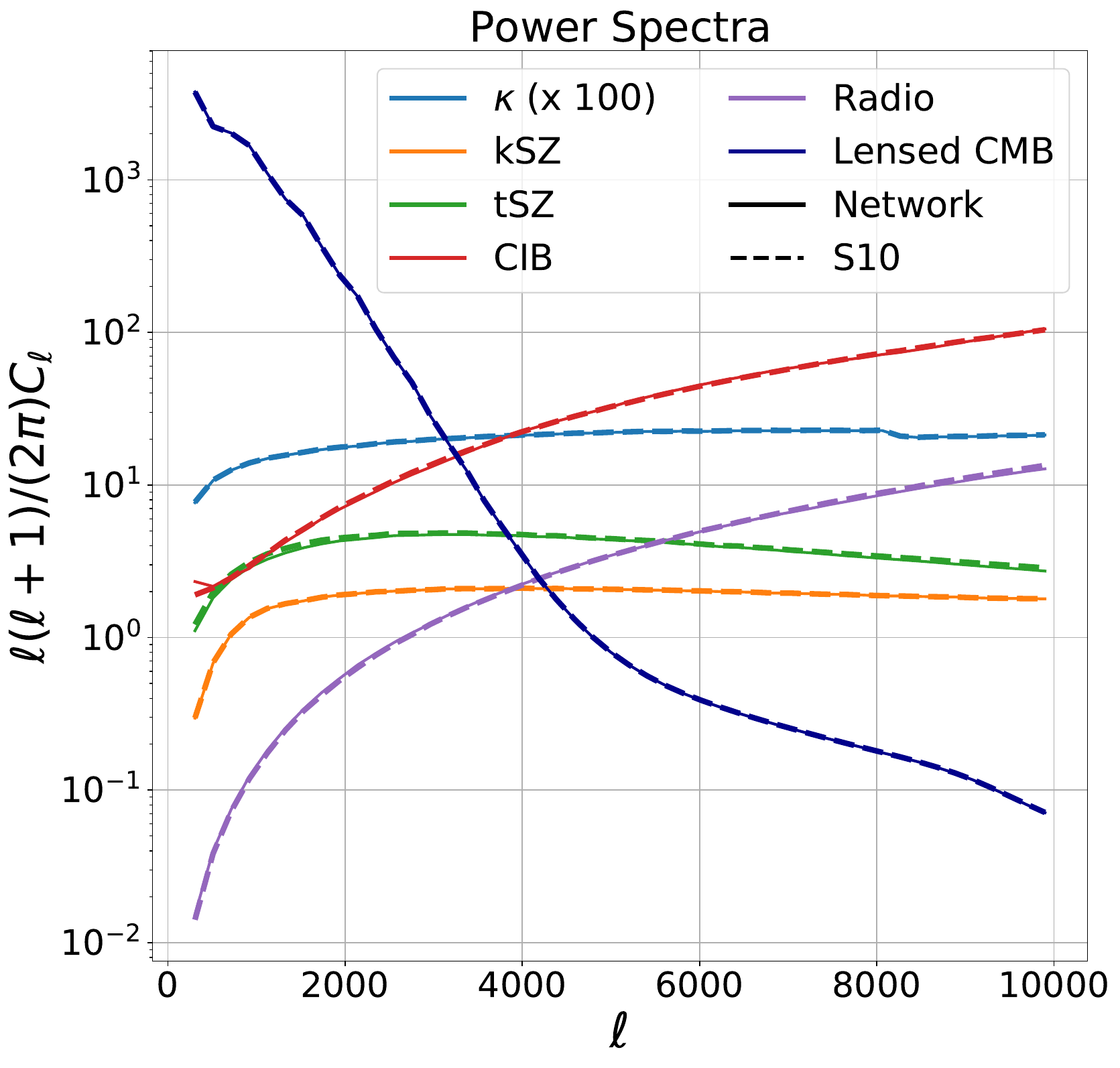}
  \caption{Shown are the power spectra of simulations at 148~GHz. The solid curves are calculated from a random sample of a full-sky simulation from the network, and dashed curves are calculated using the full-sky S10 simulation. A flux-cut of 7~mJy is applied to both CIB and Radio maps. For the purpose of visualization, we multiply the $\kappa$ map by a factor of 100 before taking the power spectrum. Since we apply one-dimensional transfer functions and scale factors, computed by comparing the network power spectra to the S10 power spectra (see Section~\ref{sec:postprocess} for details), the power spectra of the network and S10 match by construction.}
  \label{fig:ps_2pt}
\end{figure}

\begin{figure}[t]
  \centering
  \hspace{-3mm}\includegraphics[width=\columnwidth]{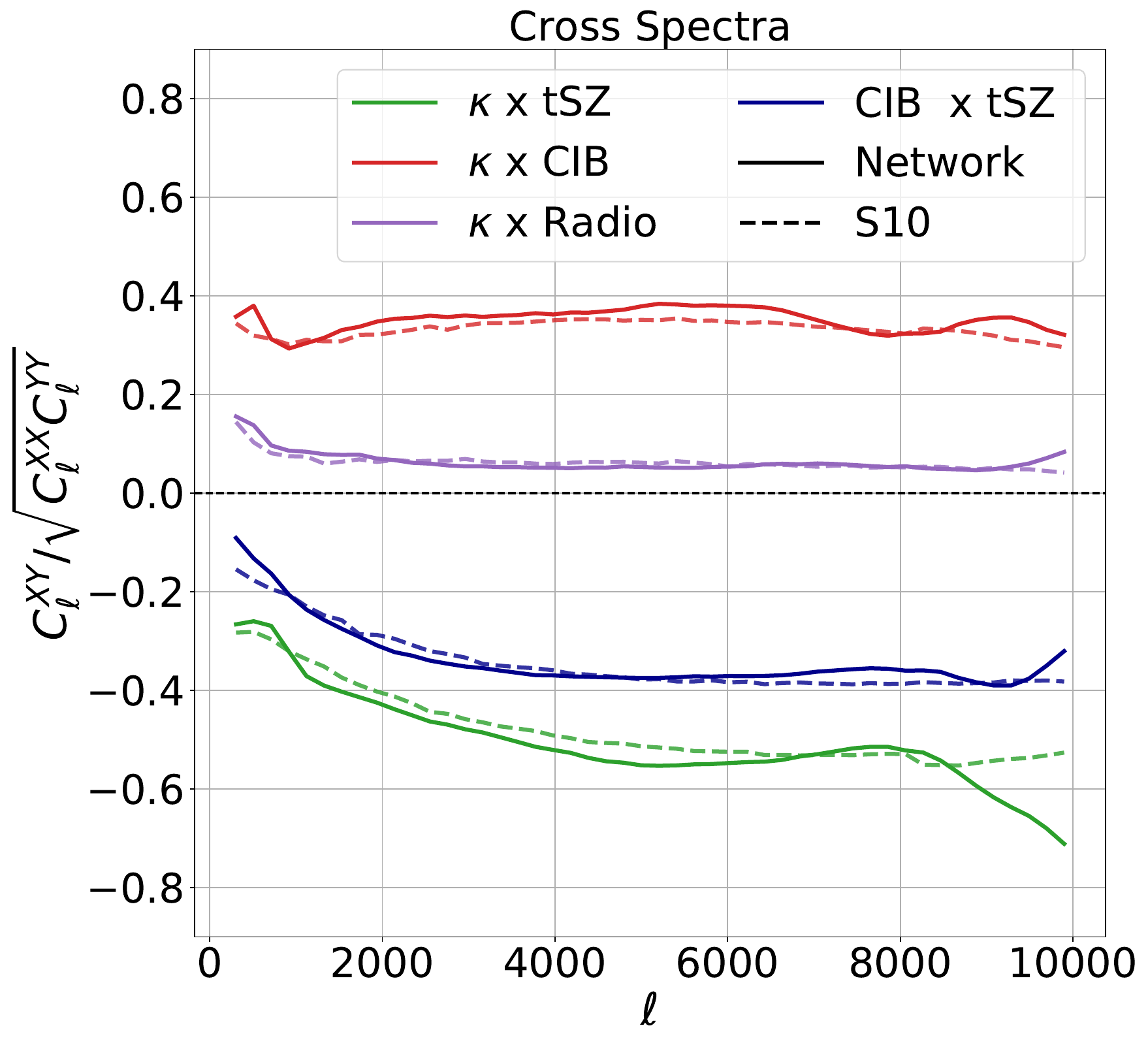}
  \caption{Shown via cross correlation coefficients are the cross spectra between the $\kappa$ map and the extragalactic foreground components at 148~GHz.  The solid curves are generated with a full-sky map from our network, while the dashed curves are generated from the S10 simulations. A flux-cut of 7~mJy is applied to both the CIB and Radio maps.  There is a good overall match between the curves. Note that we do not apply any correction to the network cross spectra; the matches between the network and the S10 cross spectra indicate that the network has correctly learned the correlations among the foreground components. } 
  \label{fig:cross_2pt}
\end{figure}

\subsection{Two-point Statistics}

Next, we show the two-point statistics of the simulations in Figure~\ref{fig:ps_2pt} and Figure~\ref{fig:cross_2pt}. Since we apply one-dimensional transfer functions and scale factors, computed by comparing the network power spectra to the S10 power spectra (see Section~\ref{sec:postprocess} for details), the power spectra of the network and S10 match by construction, as shown in Figure~\ref{fig:ps_2pt}. Figure~\ref{fig:cross_2pt} shows the cross spectra between the $\kappa$ map and the extragalactic foreground components at 148~GHz. Overall, we find a good match between the two sets of cross spectra. We find only a small extra correlation between $\kappa$ and tSZ at small scales ($\ell \ge 8000$). We also check the cross correlation among the other components (for instance the tSZ and CIB), and similarly find good matches. Note that we do not apply any correction to the network cross spectra. The matches between the network and the S10 cross spectra indicate that the network has correctly learned the correlations among the foreground components. 

\subsection{Three-point and Four-point Functions}

\begin{figure}[t]
  \centering
  \hspace{-3mm}\includegraphics[width=0.95\columnwidth]{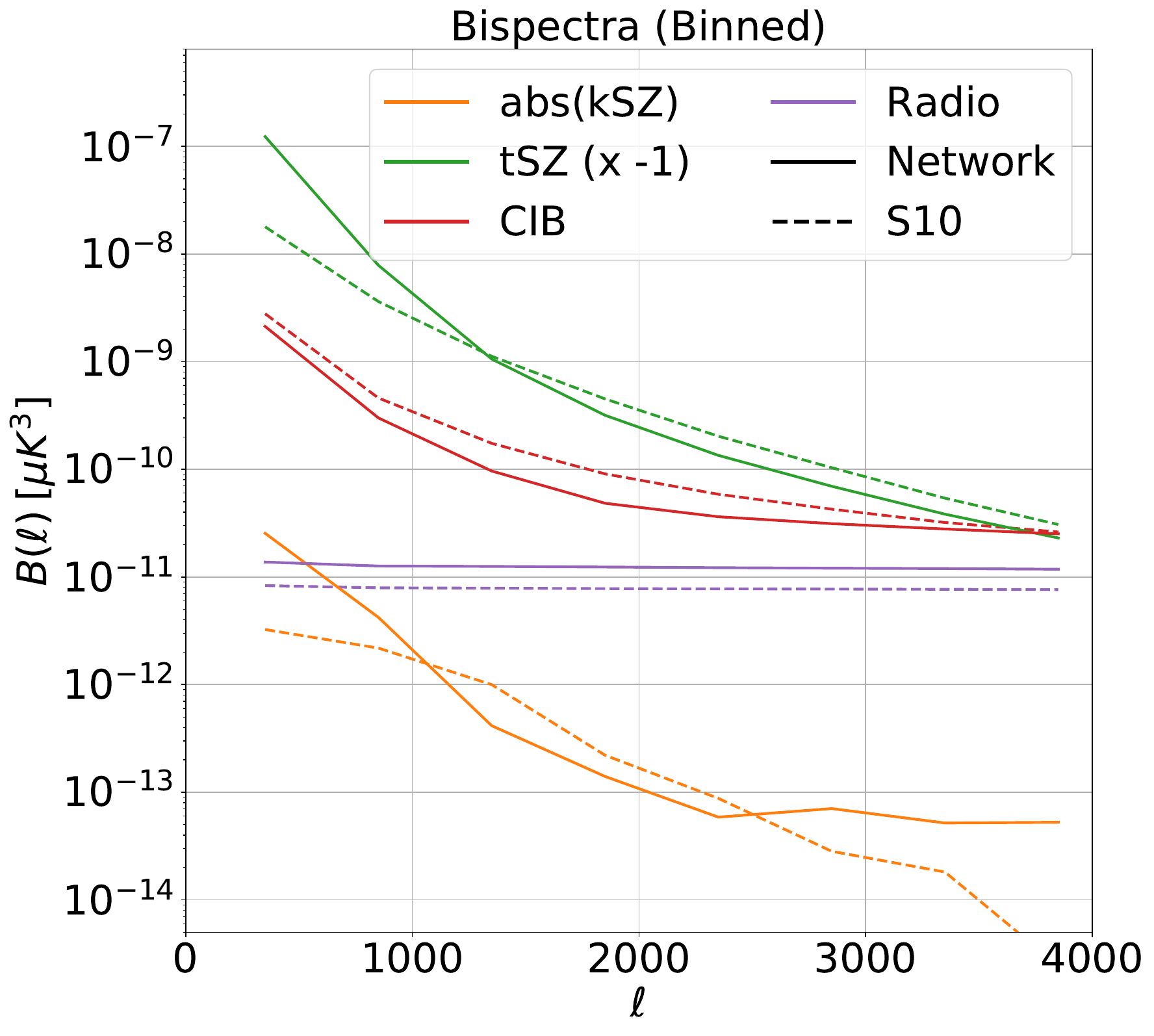}
  \caption{Shown are the binned bispectra of simulations at 148~GHz. The solid curves are calculated from a random sample of a full-sky simulation from the network, and dashed curves are calculated using the full-sky S10 simulation. A flux-cut of 7~mJy is applied to both CIB and Radio maps. Here $\ell = \sqrt{\ell_1^{2}+\ell_2^{2}+\ell_3^{2}}$, where $\ell_1$, $\ell_2$, and $\ell_3$ form a triangle. } 
  \label{fig:bispec}
\end{figure}

\begin{figure}[t]
  \centering
  \hspace{-3mm}\includegraphics[width=0.95\columnwidth]{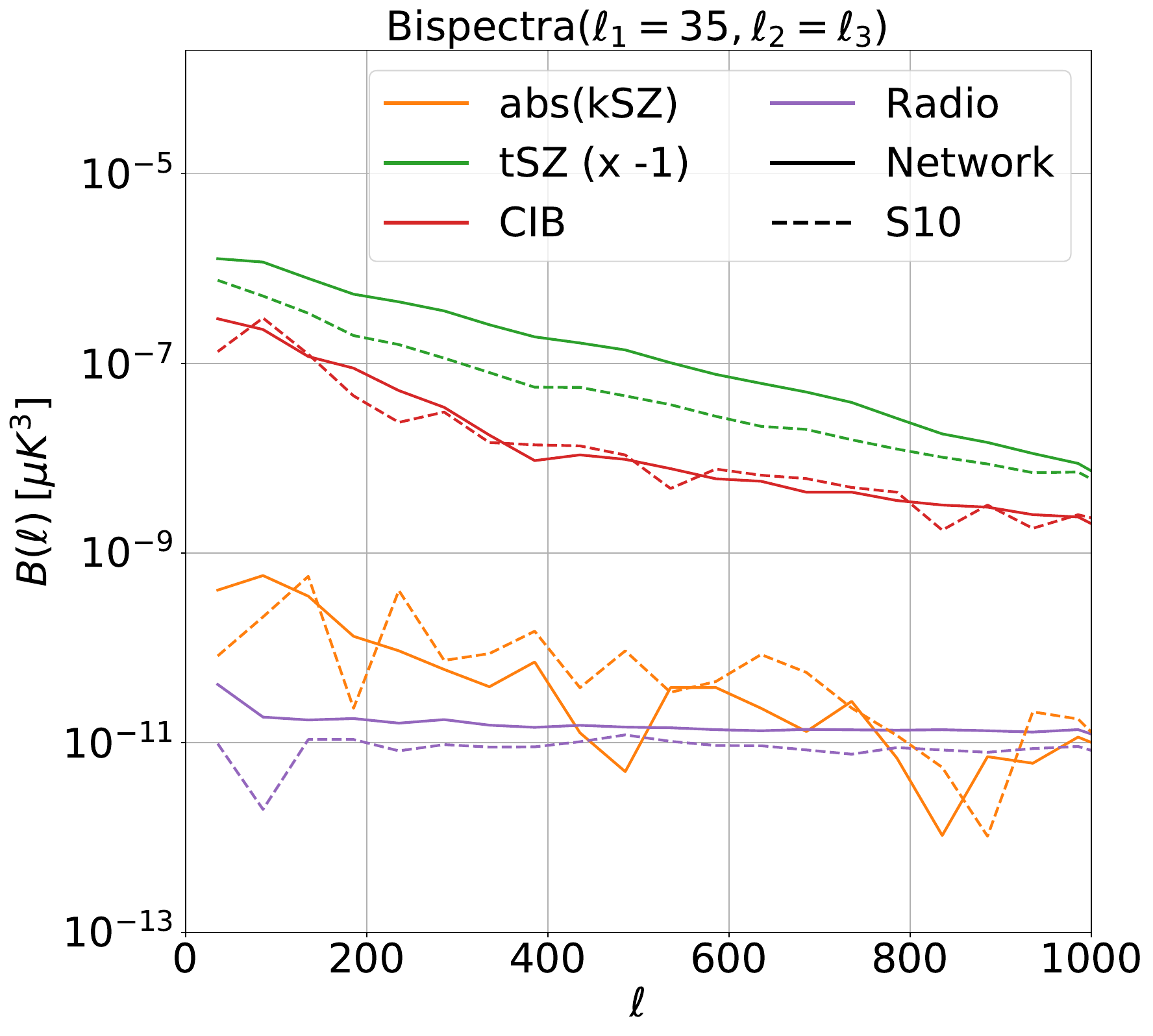}
  \caption{Similar to Figure~\ref{fig:bispec}, but for a squeezed configuration, where one leg is $\ell_1=35$, and the other two legs are the same length ($\ell_2=\ell_3)$. We find that the network can correctly reproduce the correlations between modes larger than the tile size (i.e.~$1^{\circ}$ x $1^{\circ}$ or roughly $\ell \le 200$ in the harmonic domain) and modes within the tiles (i.e.~roughly $\ell \ge 200$).}  
  \label{fig:bispec_squeeze}
\end{figure}

We also compute higher order statistics, in particular the three-point and four-point functions. We compute the bispectra of foregrounds following~\cite{Crawford2014, vanEngelen2014, Coulton2019} using the {\it PiInTheSky} software.{\color{blue}\footnote{\href{https://github.com/simonsobs/PiInTheSky}{https://github.com/simonsobs/PiInTheSky}}} The binned bispectra of the simulations are shown in Figure~\ref{fig:bispec} and Figure~\ref{fig:bispec_squeeze}. One advantage of the binned bispectrum method is that it requires no explicit modeling of the signal. 
For visualization purposes, we plot the absolute value of the kSZ bispectra, and we plot the tSZ bispectra multiplied by a factor of -1.  Overall we find a good match in the shape and amplitude of the bispectra, even though our training procedure was not explicitly optimized to reproduce the bispectra. With minor modifications to the training procedure, we expect this match can be further improved.  

In Figure~\ref{fig:bispec_squeeze},  we show a squeezed configuration bispectra, where one leg of the triangle is $\ell_1=35$, and the other two legs have the same length ($\ell_2=\ell_3$).
As discussed in Section~\ref{sec:postprocess}, we generate the full-sky realizations by "stitching" up $1^{\circ}$ x $1^{\circ}$ tiles. Naively, the network simulations are not expected to reproduce the statistics for modes larger than the tile size (i.e.~roughly $\ell \le 200$). However, we find that the network correctly reproduces the correlations between the super tile modes ($\ell_1 = 35$) and the smaller scale modes ($\ell \ge 200$). This indicates that our overall procedure can generate faithful full-sky realizations, which contain the correct information at all scales. 

\begin{figure}[t]
  \centering
  \hspace{-3mm}\includegraphics[width=\columnwidth]{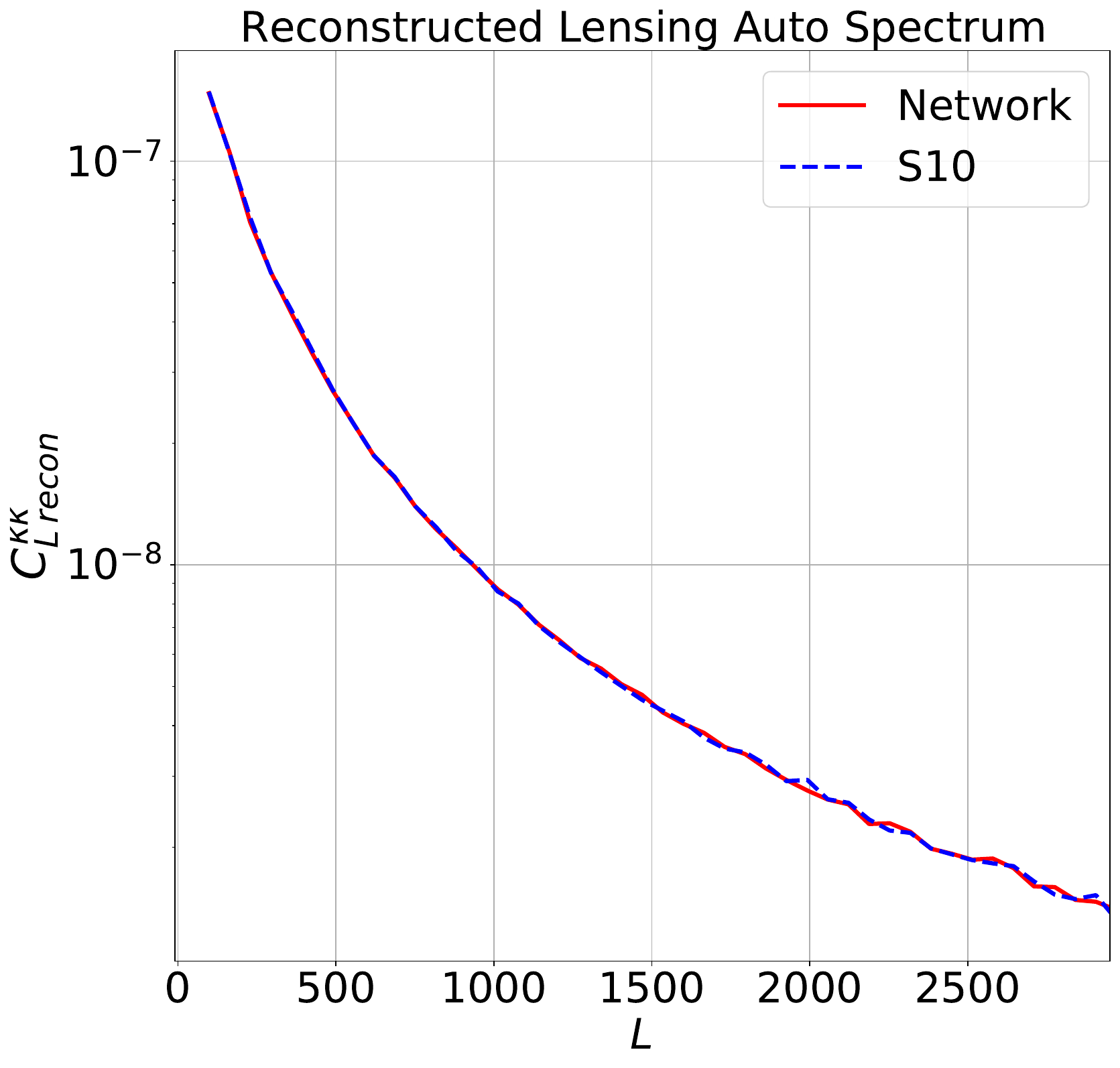}
  \caption{Shown are the reconstructed lensing auto spectra. The red solid curve is generated from a CMB map lensed with a full-sky $\kappa$ map from the network. The blue dashed curve is generated using the same CMB map, instead lensed with the full-sky S10 $\kappa$. No foregrounds are added. Lensing reconstruction is done using the temperature-only quadratic estimator, including multipoles of $\ell \in [100, 3000]$. We compute the lensing auto spectrum as the cross spectra of the reconstructed kappa map and the input kappa map in order to avoid reconstruction noise bias.}
  \label{fig:ps_4pt}
\end{figure}

\begin{figure}[t]
  \centering
  \hspace{-3mm}\includegraphics[width=\columnwidth]{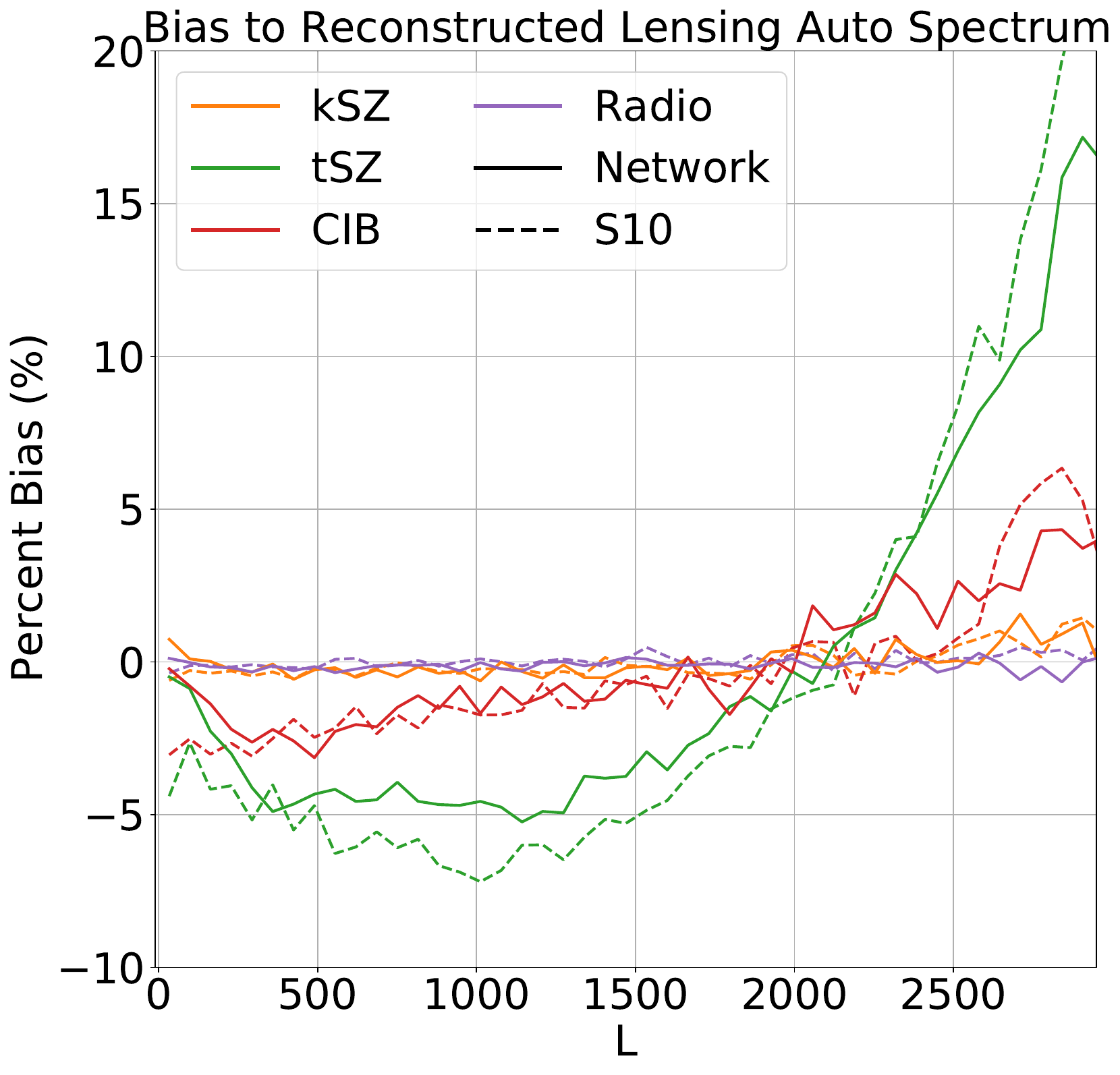}
  \caption{Shown is the percent bias to the reconstructed lensing auto spectrum induced by extragalactic foregrounds at 148~GHz. The solid curves are calculated with a random full-sky simulation from the network, while the dashed curves are computed using the S10 simulation. Lensed CMB maps are made as discussed in Figure~\ref{fig:ps_4pt}. A flux-cut of 7~mJy is applied to both CIB and Radio maps. The lensing auto spectra are computed as the cross spectra of the reconstructed and input kappa maps as in Figure~\ref{fig:ps_4pt}.  We find good agreement between the S10 simulation and the network.}
  \label{fig:bias_fgs}
\end{figure}

Figure~\ref{fig:ps_4pt} and Figure~\ref{fig:bias_fgs} show the comparison for the four-point functions used to reconstruct the lensing potential.  To do this comparison, we first generate an unlensed Gaussian CMB realization using the {\it{WMAP5}} cosmology~\cite{WMAP2009}; we then make two lensed versions of this CMB map, one lensed by the the network $\kappa$ and another lensed by the S10 $\kappa$ map.  Since our lensing reconstruction filter requires some instrument noise and a finite resolution, we add to each simulation a white noise level of 10$\mu$K-arcminute and a 1.3 arcminute Gaussian beam, which matches current ground-based CMB experiment sensitivity.  We reconstruct the lensing potential using the temperature-only quadratic estimator~\cite{Hu2002} in order to focus on the bias induced from extragalactic foregrounds.  We include CMB multipoles in the range of $\ell \in [100, 3000]$ for the reconstruction. Figure~\ref{fig:ps_4pt} shows the cross spectra between the reconstructed lensing convergence and the input lensing convergence from both the network and S10 simulations, in the absence of other foreground contamination.  We find a good match for all the scales considered.  Figure~\ref{fig:bias_fgs} shows the contamination of the reconstructed lensing convergence from simulated lensed CMB maps plus the extragalactic foregrounds as a percent bias (i.e $C_{L, biased}^{\kappa \kappa}/C_{L, unbiased}^{\kappa \kappa}\times100$, where $C_{L}^{\kappa \kappa}$ is computed from the cross spectra between the reconstructed and the input lensing convergences as described above). We find good agreement between the biases computed from the network and those from the S10 simulations. 

\section{Final Network Product}
\label{sec:product}

We make 500 mmDL realizations in total using our network, and release them publicly on the Legacy Archive for Microwave Background Data Analysis (LAMDA){\color{blue}\footnote{\href{https://lambda.gsfc.nasa.gov/simulation/tb_sim_ov.cfm}{https://lambda.gsfc.nasa.gov/simulation/tb\_sim\_ov.cfm}}} and the National Energy Research Scientific Computing Center (NERSC) cluster{\color{blue}\footnote{\href{https://crd.lbl.gov/departments/computational-science/c3/c3-research/cosmic-microwave-background/cmb-data-analysis-at-nersc/}{https://crd.lbl.gov/departments/computational-science/c3/c3-research/cosmic-microwave-background/cmb-data-analysis-at-nersc/}}}. These are full-sky simulations at half-arcminute resolution at six different frequencies (30, 90, 148, 219, 277, and 350~GHz), which include: 

\begin{itemize}
\item the lensing convergence map ($\kappa$),
\item the kinetic  Sunyaev-Zel'dovich effect (kSZ),
\item the thermal Sunyaev-Zel'dovich effect (tSZ),
\item the Cosmic Infrared Background (CIB), and
\item the radio galaxies (Radio).
\end{itemize}

For each realization, we generate six maps of the frequency-dependent components (tSZ, CIB, and Radio), and one map of the frequency-independent components (lensed CMB ($T,Q,U$), $\kappa$, and kSZ).  We release all of these components separately, as well as a combined final microwave sky map in temperature at each of the six frequencies. In total, each realization contains 29 ($6\times 3 + 5 + 6$ ) full sky maps. The simulations are natively made in the CAR pixelization and converted to HEALPix by reprojecting the CAR maps to spherical harmonic $a_{lm}$ up to an $\ell_{max}$ of 10,000 for $\kappa$, kSZ, and tSZ maps, and an $\ell_{max}$ of 25,000 for CIB and Radio maps; note that we chose a higher $\ell_{max}$ of 25,000 for CIB and Radio maps to reduce the effect of spectral leakage around the sources and to reproduce the source number counts shown in Figure~\ref{fig:source}. These $a_{lm}$ are then reprojected onto full-sky HEALPix maps of $\rm{Nside}=8192$.
We release this set of simulations in both HEALPix and CAR pixel formats, however, for space reasons we provide 500 HEALPix realizations and only 30 CAR realizations. We also provide scripts to convert between HEALPix and CAR maps. Each HEALPix fits file is 3.2~GB for double-precision, and each CAR file is 3.6~GB for single-precision.  Thus the full package is 50 TB, for $500\times29\times3.2$~GB + $30\times29\times3.6$~GB.  We also release the code used in this work at \href{https://github.com/dwhan89/cosmikyu}{https://github.com/dwhan89/cosmikyu}.

\section{Discussion}
\label{sec:discussion}

We have presented for the first time high-resolution full-sky DL simulations at millimeter wavelengths that include correlated foreground components. We find that these mmDL simulations can reproduce not only the power spectra of the original simulations, but also can naturally reproduce the other non-Gaussian statistics considered in this work (i.e.~one-point PDFs, cross-spectra, bispectra, lensing reconstructions, and lensing reconstruction biases) with good overall agreement. This is notable since we only amend the mmDL simulations to reproduce the power spectra and the source number counts above 1~mJy. 

The procedure we developed can be used to generate an unlimited number of full-sky realizations starting from just a single realization of a full-sky simulation. Ordinarily there would not be enough independent small tiles from a single full-sky simulation to train a network to predict new foreground maps given a kappa map.  However, in this work we achieve this by first generating an augmented data set (Intermediate Product 2 in Figure~\ref{fig:flowchart}) by training our network on the original simulations. This augmented data set is more than four times larger than the original one and each patch is statistically independent; thus it can be used to train the next network step (Training Step 2 in Figure~\ref{fig:flowchart}) that requires a larger amount of training data.  
This enables the capability to mass produce independent full-sky realizations from a single expensive full-sky simulation.

Currently, DL training is mostly done with small tiles due to limited computing resources (both memory and GPU). Though these resource issues will be alleviated as hardware improves overtime, we expect that they will persist for the foreseeable future. Unfortunately, these small DL simulations have a limited use when analyzing current and future wide-field CMB survey data. In this work, we outline a procedure for generating many realizations of full-sky maps, stitching together many individual tiles, that can faithfully recover the high-order statistics of a full-sky map without discontinuities or repeated features. 

An advantageous feature of our network is that the input is a full-sky kappa map (top green box in Figure~\ref{fig:flowchart}).  Thus one can now in principle take a full-sky kappa map from any large-scale structure (LSS) simulation and generate the corresponding lensed CMB and correlated foreground components at millimeter wavelengths.  This is especially useful in the current era of combining results from both CMB and LSS surveys, which require a common set of simulations.  One can, for example, input in a full-sky kappa map taken from a simulation developed for the Rubin or Roman surveys \cite{Spergel2015, LSST2019}, and generate the corresponding CMB maps, resulting in simulations with both galaxies and CMB secondary components.{\color{blue}\footnote{If the given LSS kappa map includes non-Gaussian structure, we would just remove Training Step 3 from our network.}} 

We envision these mmDL simulations useful for all parts of CMB data analyses, such as verification of pipelines, generation of covariance matrices, and analyses of foreground biases. Furthermore, since our procedure is numerically efficient, especially for generating foreground realizations on a small patch of the sky, we can use our DL method as a part of a forward modeling pipeline (such as discussed in~\cite{Millea2020}) where one varies the initial conditions to match observations.

Before we conclude, we briefly comment on the potential origin of the corrections we make in Section~\ref{sec:postprocess}. The properties of generative DL models, in particularly those of GAN, are active areas of research (see~\cite{Bau2018} for more discussion), which is out of the scope of this work. However, we discuss a couple considerations relevant to the applications of generative DL models in physics. First, there is no simple way to impose physical constraints (for instance, symmetry requirements) to neural networks, though there are studies in this direction (for example~\cite{Zhu2019, Wu2020}). This prevents DL practitioners from leveraging some physical insights important to the traditional modeling of physical processes. Second, even though the Wasserstein distance used throughout this work is the most general way to compute the distance between two distributions (i.e.~the most general discriminator), we can only compute an approximated form of the Wasserstein metric due to numerical reasons.  It is not intuitively clear whether optimizing this approximated version should always lead to physical results. These limitations are unlikely fundamental short comings of generative DL models, and we expect that they will be eventually overcome.

Finally, we note that our network is trained to the specific cosmology and baryonic model used to generate the S10 simulations. In principle, a GAN network should be able to generate simulations at different cosmologies and with different gas physics by extrapolating the latent space variables (see~\cite{Tamosiunas2020} for discussion). This is potentially a very powerful technique that can vastly expand the applicability of these mmDL simulations.

\begin{acknowledgments}
DH would like to thank the Center for Computational Astrophysics (CCA) at the Flatiron Institute for support as a CCA pre-doctoral fellow.  NS would like to thank the CCA for support during her sabbatical, during which time the idea for this project originated. DH and NS thank Julian Borrill and Andrea Zonca for their help with the product release on NERSC, and Graeme Addison for the product release on NASA LAMBDA.  DH and NS acknowledge support from NSF grant numbers AST-1513618 and AST-1907657. FVN acknowledges funding from the WFIRST program through NNG26PJ30C and NNN12AA01C. Some of the results in this paper have been derived using the HEALPix package~\cite{Gorski2005}.
\end{acknowledgments}

\appendix


\section{Normalization Parameters}
\label{sec:norm}

In the DL literature, normalization refers to a set of techniques to modify the training data in order to stabilize and to speed up the training. A few popular normalization methods are the Box-Cox transformation~\cite{Box1964} (which includes a log scaling), clipping,  min-max, and the standard score. We tried a number of different normalization methods, and found that the combination of log scaling (adopted from ~\cite{Troster2019}), and the standard score yielded the best results in our case. In Table~\ref{tab:normalization}, we list the parameter values used in our normalization procedure. These parameter values are derived from the validation data set described in Section~\ref{sec:prep2}. Each of these parameters are described in detail in Section~\ref{sec:prep3}.

\begin{table}[h]
\begin{tabular}{|c|c|c|c|}
\hline
component        & $\sigma_p$      & $\mu_{p^{'}}$ & $\sigma_{p^{'}}$ \\
\hline
$\kappa$  & N/A      & 5.342e-19   &  0.081     \\ \hline
kSZ       & N/A  &  5.173e-17    &  2.347      \\ \hline
tSZ       & 3.6809  & -0.629     & 0.286\\ \hline
CIB         & 17.134  &  0.605   &  0.349   \\ \hline
Radio         & 6.392  &  0.199     & 0.421       \\ \hline
\end{tabular}
\caption{The normalization parameters described in Section~\ref{sec:prep3}. Here $\sigma_p$ is the standard deviation of pixel values used in the log-normalization (Equation~\ref{eqn:scaled_log_norm}). $\mu_{p^{'}}$ and $\sigma_{p^{'}}$ are related to the mean and the standard deviation of the pixel values used in the standard score (Equation~\ref{eqn:standerization}). Note that the log-normalization is not applied to $\kappa$ and kSZ maps as discussed in Section~\ref{sec:prep3}.} 
\label{tab:normalization}
\end{table}

\begin{figure}[b]
  \centering
  \hspace{-3mm}\includegraphics[width=\columnwidth]{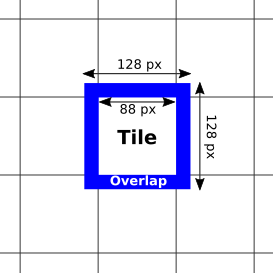}
  \caption{We show a schematic of the tiling grid described in Section~\ref{sec:postprocess}. Each tile is $128 \times 128$ pixels (roughly $1^{\circ}$ x $1^{\circ}$) in the CAR pixellization and corresponds to a patch effectively centered at the celestial equator. We cutout each tile with a 20-pixel overlap region around the edges to ensure that the final maps are continuous when we reproject the tiles back on to a full-sky CAR map. 
  }
  \label{fig:tile_layout}
  \vspace{2mm}
\end{figure}

\section{Grid schmatic}
\label{sec:grid}

Since our network can only process relatively small tiles, we need a systemic way to divide a full-sky map into a grid of tiles. For numerical reasons, the natural choice of tile width and height is $2^{n}$ pixels. In this work, we use the tile size of $128\times 128$ pixels, which is roughly $1^{\circ}$ x $1^{\circ}$. The schematic of our tile grid is shown in Figure~\ref{fig:tile_layout}. Note that we include 20 pixels overlap between tiles, which ensures continuity when we reproject tiles on to a full-sky map. This grid procedure results in a grid of tiles with 200 tiles along a longitudinal direction and 400 tiles along a latitude direction.

\section{Frequency Dependence of CIB and Radio Sources}
\label{sec:freq}

As discussed in Section~\ref{sec:postprocess}, we estimate the frequency dependence of CIB and Radio sources using the S10 catalogues. For each source in the catalogues, we compute its spectral index between a given frequency and the reference frequency of 148~GHz by comparing the flux of that source at the given frequency (30, 90, 219, 277, and 350 GHz) to the flux at the reference frequency. In Table~\ref{tab:freq}, we show the means and standard deviations of the spectral indices for CIB and Radio sources. Note that we only consider the sources that meet our flux-cut criteria ($\le \rm{7~mJy}$ at 148~GHz) in this estimation.

\begin{table}[t]
\begin{tabular}{|c|c|c|c|}
\hline
Frequency (GHz)        & $\beta_{\rm{CIB}}$      & $\beta_{\rm{Radio}}$ \\
\hline
30   & $3.304 \pm 0.138$ & $-0.799 \pm 0.158$  \\ \hline
90   & $3.202 \pm 0.442$ & $-0.811 \pm 0.144$ \\ \hline
219  & $2.973 \pm 0.567$ & $-0.820 \pm 0.135$ \\ \hline
277  & $2.870 \pm 0.372$ & $-0.822 \pm 0.134$ \\ \hline
350  & $2.745 \pm 0.301$ & $-0.823 \pm 0.133$ \\ \hline
\end{tabular}
\caption{The estimated spectral indices of CIB and Radio sources described in Section~\ref{sec:postprocess} and Appendix~\ref{sec:freq}. The spectral indices are estimated using the S10 source catalogues by comparing the flux of a source at a given frequency (30, 90, 219, 277, and 350 GHz) to the flux at the reference frequency of 148~GHz; the mean and standard deviation of the source spectral indices are listed in the Table.
}
\label{tab:freq}
\end{table}

\begin{figure}[t]
  \centering
  \hspace{-3mm}\includegraphics[width=\columnwidth]{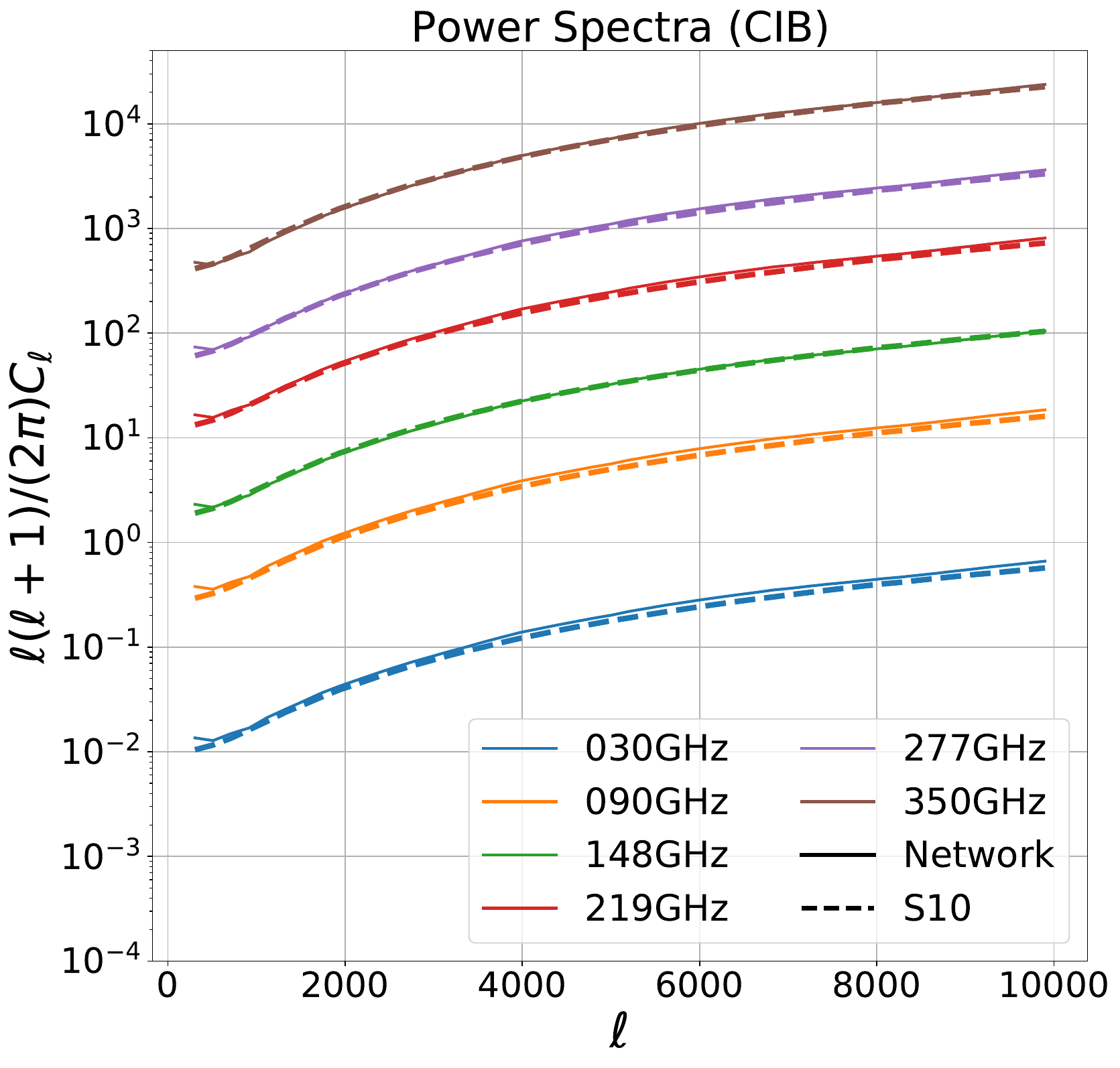}
  \caption{We show  the  power  spectra of CIB maps at different frequencies. The solid curves are calculated from a random realization of a full-sky simulation from the network, and dashed curves are calculated using the full-sky S10 simulation. A flux-cut of 7~mJy at 148~GHz is applied to all CIB maps. For each source in the 148~GHz map, we simulate its frequency dependence as described in Section~\ref{sec:postprocess} and Appendix~\ref{sec:freq}. We find good agreement for the CIB power spectra between the S10 simulation and the network.
  }
  \label{fig:cib_ps_freqs}
\end{figure}

\begin{figure}[t]
  \centering
  \hspace{-3mm}\includegraphics[width=\columnwidth]{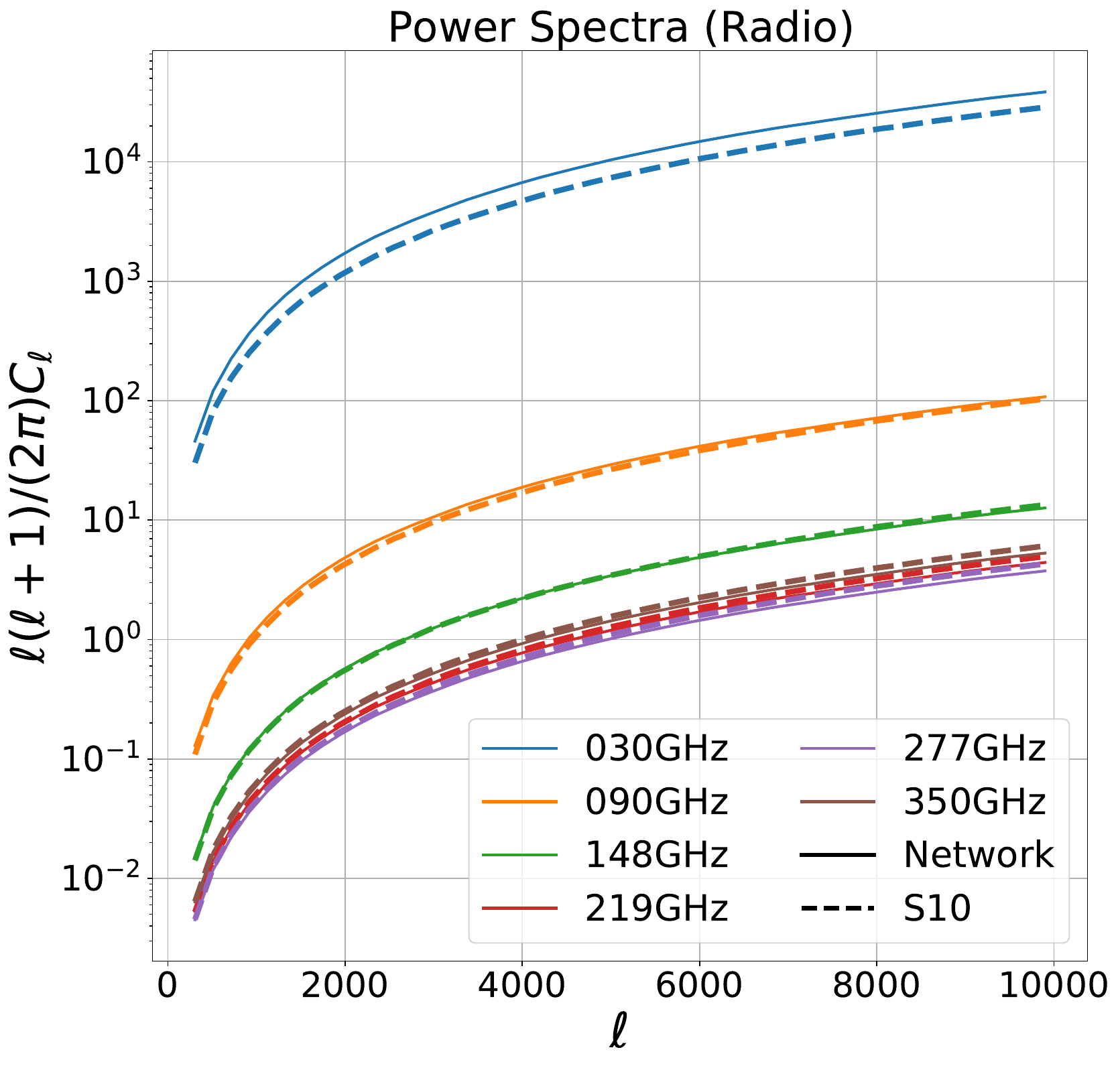}
  \caption{The same as Figure~\ref{fig:cib_ps_freqs} above, but for Radio maps instead of CIB maps.
  }
  \label{fig:radio_ps_freqs}
\end{figure}

For each source in the network 148~GHz map, we randomly assign it a spectral index for each frequency, with each index drawn from a Gaussian distribution with a mean and the standard deviation as specified in Table~\ref{tab:freq}; we then scale the source's 148~GHz flux accordingly and place it in the new frequency map. We find that we can recover well the frequency dependence of the original S10 CIB and Radio maps using this method as shown in Figure~\ref{fig:cib_ps_freqs} and Figure~\ref{fig:radio_ps_freqs}.

\section{Realization Dependent Scatter}
\label{sec:scatters}

\begin{figure}[t]
  \centering
  \hspace{-3mm}\includegraphics[width=\columnwidth]{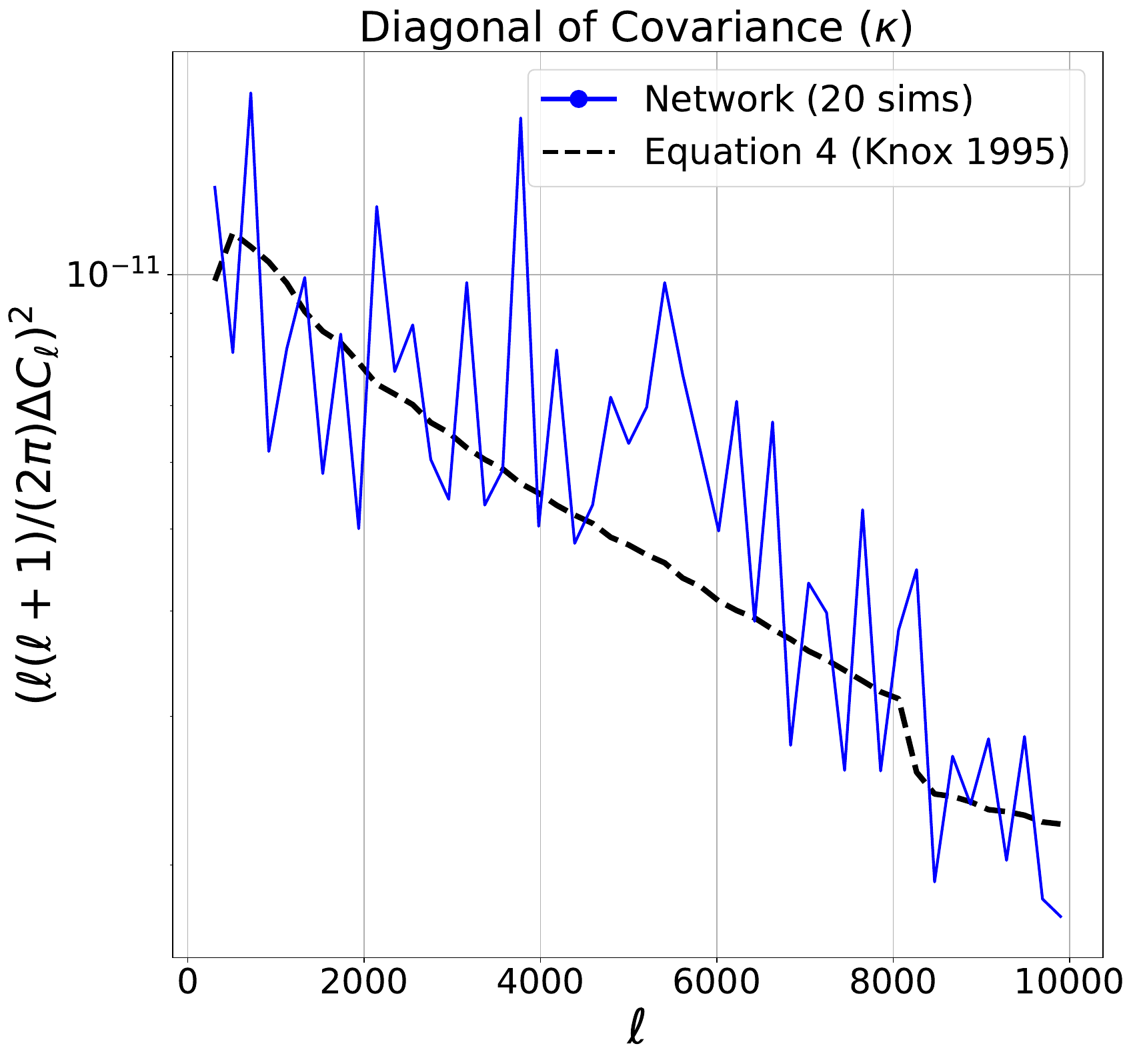}
  \caption{
  We show  the  diagonal elements of the covariance matrix derived from the $\kappa$ power spectrum. The blue curves are estimated from 20 realizations of a full-sky simulation from the network, and the dashed curve is calculated using Equation~4 in \cite{Knox1995} given the S10 $\kappa$ power spectrum.  We find good agreement between the variance from the network simulations and the expected variance. 
  }
  \label{fig:kappa_var}
\end{figure}

\begin{figure}[t]
  \centering
  \hspace{-3mm}\includegraphics[width=\columnwidth]{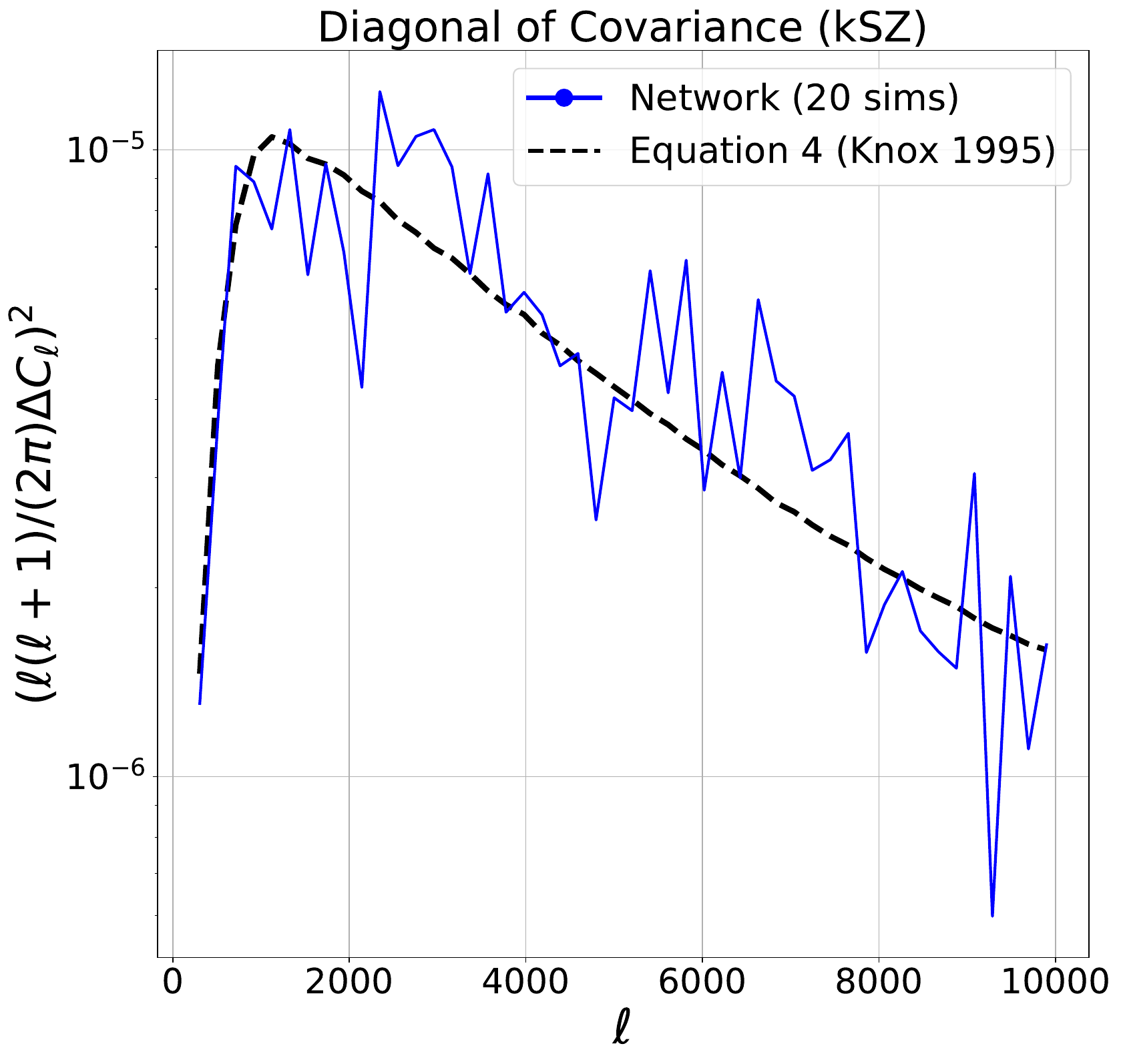}
  \caption{The same as Figure~\ref{fig:kappa_var} above, but for kSZ maps instead of $\kappa$ maps.
  }
  \label{fig:ksz_var}
\end{figure}

\begin{figure}[t]
  \centering
  \hspace{-3mm}\includegraphics[width=\columnwidth]{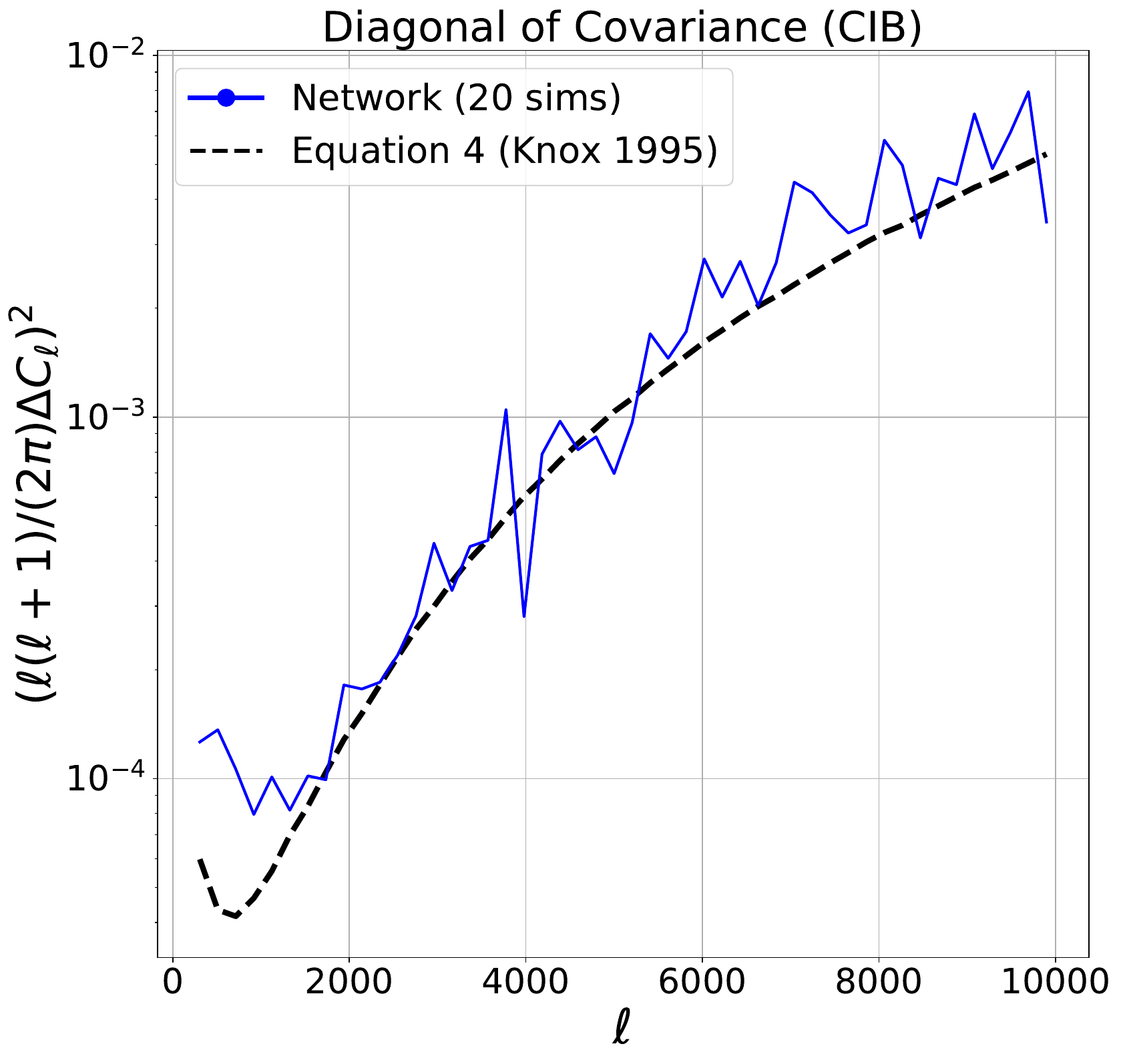}
  \caption{The same as Figure~\ref{fig:kappa_var} above, but for CIB maps at 148~GHz instead of $\kappa$ maps. Note that a flux-cut of 7~mJy at 148~GHz is applied to the CIB maps. 
  }
  \label{fig:ir_var}
\end{figure}

One common failure mode of GANs is mode collapse, which is when a GAN repeatedly generates the same or statistically similar images. In the context of our work, mode collapse will lead to a lower variance of the simulations than expected, due to the presence of repeated tiles. In this section, we empirically show that the variances of our network simulations match the expected variances. Figure~\ref{fig:kappa_var} shows the diagonal elements of the covariance matrix derived from the $\kappa$ power spectrum of 20 network simulations. Here, the expected variance is given by Equation~4 in \cite{Knox1995}; in the absence of experimental noise, this equation gives the expected variance of an approximately Gaussian random field. Similarly, we show the estimated variance of 20 network kSZ and CIB maps in Figure~\ref{fig:ksz_var} and Figure~\ref{fig:ir_var}.
For each of these components, we find that the variance from the network simulations matches the expected variance. Since each lensed CMB realization is generated from independent unlensed CMB realizations, the lensed CMB maps should have the correct variance by the construction.

\section{Supplementary Network Training Details}
\label{sec:training}

The details of our network architectures are discussed in Section~\ref{sec:method}. In this section, we provide supplementary details of our training procedure. Table~\ref{tab:training} summarizes a few additional parameters of our network training procedure. As shown in the table, all three GAN layers share most of the common parameters including the learning rate and the Adam optimizer parameters. We check our metrics (images, loss function, power spectra, and others) every 100,000 training data samples processed (i.e.~two times per epoch for DCWGAN-GP and VAEGAN; eight times per epoch for PIXGAN). If we find a degradation of the output quality, we restart the training from the last epoch with the learning rate reduced by a factor of two. Also, if the model either can not reproduce the power spectra after the loss function plateaus or images are not visually correct, we start the training from scratch with the learning rate reduced by a factor of two. Lastly, we do not explicitly check for mode collapse, but rely on the summary statistics (shown in Section~\ref{sec:results}) and the estimated variances (shown in Appendix~\ref{sec:scatters}) to infer the presence of mode collapse.

\begin{table}
\centering
\begin{tabular}{|l|c|c|c|} 
\hline
                     & DCWGAN-GP            & PIXGAN               & VAEGAN                    \\ 
\hline
Type                 & Non Conditional      & Conditional          & Conditional               \\ 
\hline
Loss function        & \multicolumn{3}{c|}{Wasserstein}                                        \\ 
\hline
Optimize             & \multicolumn{3}{c|}{Adam Optimizer ($\beta_1=0.5$, $\beta_2=0.9$)}  \\ 
\hline
Learning Rate        & \multicolumn{3}{c|}{Initially $lr=10^{-4}$. Halved if needed}    \\ 
\hline
Batch Size           & \multicolumn{3}{c|}{32}                                                 \\ 
\hline
\multicolumn{1}{l}{} & \multicolumn{1}{l}{} & \multicolumn{1}{l}{} & \multicolumn{1}{l}{}     
\end{tabular}
\caption{Shown are additional parameters used to train the three GAN layers discussed in Section~\ref{sec:method}. 
}\label{tab:training}
\end{table}

\clearpage
\bibliography{main.bib}

\end{document}